\documentclass[10pt,twocolumn,twoside,journal]{IEEEtran}
\usepackage{amsmath,graphicx}

\usepackage{multirow}
\usepackage{algorithm}
\usepackage{algorithmic}
\usepackage{amsmath}
\usepackage{amssymb}
\usepackage{bbm}
\usepackage{bm}
\usepackage{color}
\usepackage{dsfont}
\usepackage{graphicx}
\usepackage{times}
\usepackage{caption}
\usepackage{subcaption}
\usepackage{tikz}
\usepackage{url}
\usetikzlibrary{arrows,shapes,snakes,automata,backgrounds,petri}
\tikzstyle{observation} = [shading=axis, shading angle=10]
\usepackage{epsfig}
\usepackage{verbatim}

\newcommand{\commentout}[1]{}
\newcommand{\junk}[1]{}

\newcommand{\numpath}{r}

\def\piedit#1{{\color{violet}[\bf PI:~#1]}}

\def\revision#1{{\color{black} #1}}

\usepackage{cite}

\ifCLASSINFOpdf
\else
\fi
\usepackage{array}
\begin{document}
%
\title{Node Embedding via Word Embedding for Network Community
  Discovery}
%
%
%
%

\author{Weicong~Ding,~\IEEEmembership{}
        Christy~Lin,~\IEEEmembership{}
        and~Prakash~Ishwar,~\IEEEmembership{Senior~Member,~IEEE}
\thanks{W.~Ding is with Amazon, Seattle, USA,
  e-mail: weicding@amazon.com.}%
\thanks{C.~Lin is with the Division of Systems Engineering, Boston
  University, Boston, MA, 02215 USA e-mail: cy93lin@bu.edu.}%
\thanks{P.~Ishwar is with the Department of Electrical and Computer
  Engineering, Boston University, Boston, MA, 02215 USA e-mail:
  pi@bu.edu.}
}

\maketitle

\begin{abstract}
Neural node embeddings have recently emerged as a powerful
representation for supervised learning tasks involving
graph-structured data. We leverage this recent advance to develop
a novel algorithm for unsupervised community discovery in graphs.
Through extensive experimental studies on simulated and real-world
data, we demonstrate that the proposed approach consistently improves
over the current state-of-the-art. Specifically, our approach
empirically attains the information-theoretic limits for community
recovery under the benchmark Stochastic Block Models for graph
generation and exhibits better stability and accuracy over both
Spectral Clustering and Acyclic Belief Propagation in the community
recovery limits.
\end{abstract}


%
\IEEEpeerreviewmaketitle

\section{Introduction}
\label{sec:intro}

\IEEEPARstart{L}{earning} a representation for nodes in a graph, also
known as node embedding, has been an important tool for extracting
features that can be used in machine learning problems involving
graph-structured data \cite{bengio2013representation,
  belkin2002laplacian, deepwalk15, yang2016revisiting}.
Perhaps the most widely adopted node embedding is the one based on the
eigendecomposition of the adjacency matrix or the graph Laplacian
\cite{belkin2002laplacian, SC4Mc, SC1large}.
Recent advances in word embeddings for natural language processing
such as \cite{mikolov2013distributed} has inspired the development of analogous
embeddings for nodes in graphs
\cite{node2vec16, deepwalk15}.
These so-called ``neural'' node embeddings have been applied to a
number of supervised learning problems such us link prediction and
node classification and demonstrated state-of-the-art performance
\cite{node2vec16, deepwalk15, yang2016revisiting}.

In contrast to applications to supervised learning problems in graphs,
in this work we leverage the neural embedding framework to develop an
algorithm for the {\it unsupervised} community discovery problem in
graphs \cite{NMI, fortunato2010community, abbe2015, yang2015defining}.
The key idea is straightforward: learn node embeddings such that
vectors of similar nodes are close to each other in the latent
embedding space. Then, the problem of discovering communities in a
graph can be solved by finding clusters in the embedding space.
We focus on non-overlapping communities and validate the performance
of the new approach through a comprehensive set of experiments on both
synthetic and real-world data. Results demonstrate that the
performance of the new method is consistently superior to those of
spectral methods across a wide range of graph sparsity levels.
%
In fact, we find that the proposed algorithm can empirically attain
the {\it information-theoretic phase transition} thresholds for exact
and weak recovery of communities under the Stochastic Block Model
(SBM)~\cite{BP14, decelle2011KS, abbe2015, abbe2016nips}.
SBM is a canonical probabilistic model for random graphs with latent
structure and has been widely used for empirical validation and
theoretical analysis of community detection algorithms
\cite{white1976social, SBMapp, fortunato2010community, NMI}. In
particular, when compared to the best known algorithms based on
Acyclic Belief Propagation (ABP) that can provably detect communities
at the information-theoretic limits \cite{abbe2016nips, abbe2015,
  decelle2011KS}, our approach has consistently better accuracy. In
addition, we find that ABP is very sensitive to random initialization
and exhibits high variability. In contrast, our approach is stable to
both random initialization and a wide range of algorithm parameter
settings.

Our implementation and scripts to recreate all the results in this paper are available at \url{https://github.com/cy93lin/SBM_node_embedding}


\section{Related Works}

%
The community detection problem has been extensively studied in the
literature \cite{NMI, fortunato2010community, yang2015defining,
  leskovec2008statistical},\revision{\cite{shuman2013emerging}}. It
has important applications in various real-world networks that are
encountered in sociology, biology, \revision{signal processing},
statistics and computer science.
One way to systematically evaluate the performance of a community
detection algorithm and establish theoretical guarantees is to
consider a generative model for graphs with a latent community
structure.
The most widely-adopted model is the classic Stochastic Block Model.
SBM was first proposed in \cite{white1976social, SBMoriginal,
  boppana1987eigenvalues} as a canonical model for studying community
structure in networks and various community detection algorithms based
on it have been proposed, e.g., \cite{SC3LR, SC4Mc, SC2RCY,
  hajek2016achieving}.
\revision{Among these approaches, algorithms that are based on the
  graph spectrum and semidefinite programming relaxations of suitable
  graph-cut objectives have been extensively
  studied~\cite{SC2RCY,hajek2016achieving}. In particular, the phase
  transition behavior of spectral graph clustering for a generative
  model that includes SBM as a special case has also been established
  recently~\cite{chen2016phase}.
%
Graph-statistics based algorithms such as modularity optimization and
their connection to Bayesian models on graphs have also been studied
\cite{newman2016community}}.
Only very recently have the information-theoretic limits for community
recovery under the general SBM model been established \cite{BP14,
  abbe2015, abbe2016nips}.
In \cite{abbe2015, abbe2016nips}, a belief-propagation based algorithm
has been shown to asymptotically detect the latent communities in an
SBM and achieve the information-theoretic limits.
It has also been shown that graph spectrum based algorithms cannot
achieve the information-theoretic limits for recovering communities in
SBM models \cite{abbe2015}.

The use of a neural-network to embed natural language words into
Euclidean space was made popular by the famous ``word2vec''
algorithm~\cite{mikolov2013distributed, pennington2014glove}.
In these works, each word in a vocabulary is represented as a
low-dimensional vector in Euclidean space. These representations are
learned in an unsupervised fashion using large text corpora such as
Wikipedia articles.
\revision{The neural word embedding idea was adapted in
  \cite{deepwalk15} to embed nodes from a graph into Euclidean space
  and use the node embedding vectors to solve supervised learning
  tasks such as node attribute prediction and link prediction. This
  method has also been used in \cite{yang2016revisiting} to solve
  semi-supervised learning tasks associated with graphs. The node
  embeddings are computed by viewing nodes as ``words'', forming
  ``sentences'' via random paths on the graph, and then employing a
  suitable neural embedding technique for words. Different ways of
  creating ``sentences'' of nodes was further explored in
  \cite{node2vec16} where a parametric family of node transition
  probabilities was proposed to generate the random paths. The
  transition probabilities need node and/or edge labels and is
  therefore only suitable for supervised tasks.}

\revision{Our work is most closely related to \cite{deepwalk15,
    node2vec16}. While \cite{deepwalk15, node2vec16} make use of node
  embeddings in supervised learning problems such as node attribute
  prediction and link prediction, this paper focuses on the
  unsupervised community detection problem. We also explore the
  information-theoretic limits for community recovery under the
  classic SBM generative model and empirically show that our algorithm
  can achieve these limits.}

\revision{Random walks have been used in a number of ways to detect
  communities. Seminal work in \cite{spielman2004nearly} and
  \cite{andersen2006local} proposed to use a random walk and its
  steady-state-distribution for graph clustering. Subsequent work
  \cite{lambiotte2014random} further proposed to exploit multi-step
  transition probabilities between nodes for clustering. Our work can
  be viewed as implicitly factorizing a gram-matrix related to the
  multi-step transition probabilities between nodes ({\it
    cf.}~Sec.~\ref{sec:alg}). This is different from the prior
  literature. The idea of converting a graph into a time-series signal
  or a time-series signal into a graph has also been studied in the
  signal processing community and applied to the problem of graph
  filtering~\cite{hamon2015graphs}.}


\revision{Our algorithm is also related to graph clustering methods
  that leverage local connectivity structure through, for example, the
  graph wavelet transform or a suitable graph convolution operator
  \cite{tremblay2014graph,kipf2016semi,shuman2013emerging}.}
\section{Node Embedding for Community Discovery}
\label{sec:alg}
Let $\mathcal{G}$ be a graph with $n$ nodes and $K$ latent
communities. We focus on non-overlapping communities and denote by
$\pi_{i}\in\{1,\ldots, K\}$ the latent community assignment for node
$i$. Given $\mathcal{G}$, the goal is to infer the community
assignment $\hat{\pi_{i}}$.

Our approach is to learn, in an unsupervised fashion, a
low-dimensional vector representation for each node that captures its
local neighborhood structure. These vectors are referred to as {\it
  node embeddings}.  The premise is that if done correctly, nodes from
the same community will be close to each other in the embedding
space. Then, communities can be found via clustering of the node
embeddings.

\revision{\noindent{\bf Skip-gram word-embedding framework:}} In order
to construct the node embedding, we proceed as in the skip-gram-based
negative sampling framework for word embedding which was recently
developed in the natural language processing literature
\cite{mikolov2013distributed, deepwalk15}.
A document is an ordered sequence of words from a fixed vocabulary. A
$w$-skip-bigram is an ordered pair of words $(i,j)$ that occur within
a distance of $w$ words from each other 
within a sentence in the document.
%
%
A document is then viewed as a multiset $\mathcal{D}_+$ of all its
$w$-skip-bigrams $(i,j)$ which are generated in an independent and
identically distributed (IID) fashion according to a {\it joint}
probability $p((i,j))$ which is related to the word embedding vectors
$\mathbf{u}_i, \mathbf{u}_j \in \mathbb{R}^{d}$, of 
words $i$ and $j$ respectively, in $d$-dimensional Euclidean space.

Now consider a multiset $\mathcal{D}_-$ of $w$-skip-bigrams $(i,j)$
which are generated in an IID fashion according to the {\it product}
probability $p((i))\cdot p((j))$ where the $p((i))$'s are the unigram
(single word) probabilities. The unigram probabilities can be
approximated via the empirical frequencies of individual words
(unigrams) in the document.

The $w$-skip-bigrams in $\mathcal{D}_+$ are labeled as positive
samples ($D = +1$) and those in $\mathcal{D}_-$ are labeled as
negative samples ($D=-1$). In the negative sampling framework
\cite{mikolov2013distributed, deepwalk15}, the {\it posterior} probability that an
observed $w$-skip-bigram will be labeled as positive is modeled as
follows
\begin{eqnarray}
p(D = +1 | (i,j)) = 1 - p(D = -1 | (i,j)) = \frac{1}{1+
  e^{-\mathbf{u}_{i}^{\top}\mathbf{u}_{j}}}
\label{eq:softmax}
\end{eqnarray}
Under this model, the likelihood ratio $p((i,j)|D =
+1)/p((i,j)|D=-1)$, becomes proportional to
$e^{\mathbf{u}_{i}^{\top}\mathbf{u}_{j}}$.  Thus the negative sampling
model posits that the ratio of the odds of observing a $w$-skip-bigram
from a bonafide document to the odds of observing it due to pure
chance is exponentially related to the inner product of the underlying
embedding vectors of the nodes in the $w$-skip-bigram.

\revision{\noindent{\bf Maximum-likelihood estimation of embeddings:}} The word
embedding vectors $\{{\bf u}_i\}$ which are parameters of the
posterior distributions are selected to maximize the posterior
likelihood of observing all the positive and negative samples, i.e.,
\begin{eqnarray*}
\arg\max_{\mathbf{u}_{i}} \prod_{(i,j)\in \mathcal{D}_+}p(D=+1|(i,j))
\prod_{(i,j)\in \mathcal{D}_-}p(D=-1|(i,j))
\end{eqnarray*}
Substituting from Eq.~(\ref{eq:softmax}) and taking negative log, this
reduces to
\begin{eqnarray}
\arg\min_{\mathbf{u}_{i}} \left[\sum_{(i,j)\in
    \mathcal{D}_+}\log(1+e^{-\mathbf{u}_{i}^{\top}\mathbf{u}_{j}}) +
  \sum_{(i,j)\in
    \mathcal{D}_-}\log(1+e^{+\mathbf{u}_{i}^{\top}\mathbf{u}_{j}})\right]
\label{eq:w2v_negsam}
\end{eqnarray}
\revision{The optimization problem in Eq.~(\ref{eq:w2v_negsam}) can be
  reformulated as
\begin{eqnarray}
\arg\min_{\mathbf{u}_{i}}
\sum_{(i,j)}\left[n_{ij}^+\log(1+e^{-\mathbf{u}_{i}^{\top}\mathbf{u}_{j}})
  + n_{ij}^{-}\log(1+e^{+\mathbf{u}_{i}^{\top}\mathbf{u}_{j}})\right]
\label{eq:w2v_negsam2}
\end{eqnarray}
where the summation is over all distinct pairs of words $(i,j)$ in the
vocabulary and $n_{ij}^{+}$ and $n_{ij}^{-}$ are the number of $(i,j)$
pairs in $\mathcal{D}_{+}$ and $\mathcal{D}_{-}$ respectively. The
objective function in Eq.~(\ref{eq:w2v_negsam}) or equivalently
Eq.~(\ref{eq:w2v_negsam2}) is non-convex with respect to the embedding
vectors. Moreover, the solution is not unique because the objective
function, which only depends on the pairwise inner products of the
embedding vectors, is invariant to any global angle-preserving
transformation of the embedding vectors.}

\revision{One solution approach \cite{levy2014neural} is to first
  re-parameterize the objective in terms the of the gram matrix of
  embedding vectors $G$, i.e., replace
  $\mathbf{u}_{i}^{\top}\mathbf{u}_{j}$ with $G_{ij}$, the $ij$-th
  entry of $G$, and then solve for the optimum $G$ by relaxing the
  requirement that $G$ is symmetric and positive semi-definite. The
  solution to this relaxed problem is given by $G_{ij} =
  \log(n_{ij}^+/n_{ij}^-)$ which can be shown to be equal, up to an
  additive constant, to the so-called pointwise mutual information
  (PMI) $\log(\#(i,j)/ (\#(i) \cdot \#(j) ))$, where $\#$ denotes the
  number of occurrences in $\mathcal{D}_+$. The embedding vectors can
  then be obtained by performing a low-rank matrix factorization of
  $G$ via, for example, an SVD.}

\revision{An alternative solution approach which we adopt in our
  node-embedding algorithm described below, is to optimize
  Eq.~(\ref{eq:w2v_negsam}) using stochastic gradient descent (SGD)
  \cite{bottou2010large,bottou-98x}. SGD iteratively updates the
  embedding vectors by moving them along directions of negative
  gradients of a modified objective function which is constructed
  (during each iteration) by partially summing over a small, randomly
  selected, batch of terms that appear in the complete summation that
  defines the original objective function
  (cf.~Eq.~(\ref{eq:w2v_negsam})). This is the approach that is
  followed in \cite{mikolov2013distributed}. An advantage of SGD is
  its conceptual simplicity. The other advantage is that it can be
  parallelized and nicely scaled to large
  vocabularies~\cite{recht2011hogwild}. SGD also comes with
  theoretical guarantees of almost sure convergence to a local minimum
  under suitable regularity conditions \cite{bottou-98x}.}

\revision{
\noindent{\bf Proposed node-embedding algorithm:} We convert the word
embedding framework for documents described above into a node
embedding framework for graphs. Our key idea} is to view nodes as
words and and a document as a collection of sentences that correspond
to paths of nodes in the graph.
%
%
To operationalize this idea, we generate multiple paths (sentences)
%
%
by performing random walks of suitable lengths starting from each
node.
Specifically, we simulate $\numpath$ random walks on $\mathcal{G}$ of
fixed length $\ell$ starting from each node. \revision{In each random
  walk, the next node is chosen uniformly at random among all the
  immediate neighbors of the current node in the given graph.} The set
$\mathcal{D}_+$ is then taken to be the multiset of all node pairs
$(i,j)$ for each node $i$ and all nodes $j$ that are within $\pm w$
steps of node $i$ in all the simulated paths {\it whenever} $i$
appears. \revision{The parameter $w$ controls the size of the local
  neighborhood of a node in the given graph. The local neighborhood of
  a node is the counterpart of {\it context words} surrounding a word
  in a given text document.}
%
%

The set $\mathcal{D}_-$ (negative samples) is constructed as a
multiset using the following approach: for each node pair $(i,j)$ in
$\mathcal{D}_+$, we append $m$ node pairs $(i, j_1), \ldots, (i, j_m)$
to $\mathcal{D}_-$, where the $m$ nodes $j_1,\ldots,j_m$ are drawn in
an IID manner from {\it all} the nodes according to the estimated
unigram node (word) distribution across the document of node paths.
%
%
\revision{The set $\mathcal{D}_-$ captures the behavior of a random
  walk on a graph which is completely connected. When applied to
  graphs as we do, the negative sampling model can be viewed as
  positing that the ratio of the odds of observing a pair of nodes
  that are within $w$ steps from each other in a random walk on the
  {\it given graph} to the odds of observing the same pair in a random
  walk on a (suitably edge-weighted) {\it completely connected graph}
  is exponentially related to the inner product of the underlying
  embedding vectors of the pair of nodes.}

Once $\mathcal{D}_+$ and $\mathcal{D}_-$ are generated, we optimize
Eq.~(\ref{eq:w2v_negsam}) using stochastic gradient descent
\cite{mikolov2013distributed}. \revision{The per-iteration
  computational complexity of the SGD algorithm used to solve
  Eq.~(\ref{eq:w2v_negsam}) is $O(d)$, i.e., linear in the emebdding
  dimension. The number of iterations is $O(mrlw)$.}

Once the embedding vectors $\mathbf{u}_i$'s are
learned, we apply $K$-means clustering to get the community
memberships for each node. These steps are summarized in
Algorithm~\ref{alg}.
\begin{algorithm}[!htb]
\caption{VEC: Community Discovery via Node \\ Embedding}
\label{alg} 
\begin{algorithmic}
\STATE {\bf Input:} Graph $\mathcal{G}$, Number of communities $K$;
Paths per node $\numpath$, Length of path $\ell$, Embedding dimension
$d$, Contextual window size $w$
\STATE {\bf Output:} Estimated Community memberships
$\hat{\pi}_1,\ldots, \hat{\pi}_n$
\FOR {Each node $v$ and $ t \in \{1 \dots \numpath\}$ }
	\STATE $\mathbf{s}_{v,t} \leftarrow$ A random path of length
        $\ell$ starting from node $i$
%
\ENDFOR
\STATE $\{\hat{\mathbf{u}}_{i}\}_{i=1}^{n} \leftarrow $ solve
Eq.~(\ref{eq:w2v_negsam}) with paths $\{\mathbf{s}_{v,t}\}$ and window
size $w$.
\STATE $\hat{\pi}_1,\ldots, \hat{\pi}_n \leftarrow $
$K$-means$(\{\hat{\mathbf{u}}_{i=1}^{n}\}_{i}, K )$
\end{algorithmic}
\end{algorithm}

\revision{\noindent{\bf Selecting algorithm parameters}: The proposed
  algorithm VEC has 5 tuning parameters. These are (i) $r$: the number
  of random walks launched from each node, (ii) $l$: length of each
  random walk, (iii) $w$ the local window size, (iv) $d$: the
  embedding dimension, and (v) $m$: the number of negative samples per
  positive sample. In general, the community recovery performance of
  VEC will depend on all $5$ of these tuning parameters.  We do not
  currently have a rigorous theoretical framework which can guide the
  optimum joint selection of all these parameters. An exhaustive
  exploration of the five-dimensional space of all algorithm
  parameters to determine which combinations of choices have good
  performance for graphs of different sizes (number of nodes),
  different sparsities (number of edges), and number of communities is
  clearly impractical.}

  \revision{In Sec.~\ref{sec:param-sensitive} we explore the
    sensitivity of the community recovery performance of VEC to
    perturbations of algorithm parameters around the following default
    setting: $\numpath =10$, $\ell=60$, $w = 8$, and $d=50$.  We set
    the number of negative samples per observation to five ($m = 5$)
    as suggested in \cite{mikolov2013distributed}.
  The results from Sec.~\ref{sec:param-sensitive} demonstrate that the
  performance of VEC remains remarkably stable across a wide range of
  values of algorithm-parameters around the default setting.
  The only two significant parameters that seem to have a noticeable
  impact on community recovery performance are $w$ and, to a lesser
  extent, $d$. Informally, we can try to make sense of these empirical
  observations as follows. If the graph is fully connected and
  aperiodic, then as $l \rightarrow \infty$, the node distribution
  will converge to the stationary distribution of the Markov chain
  defined by the graph. It is therefore not surprising that the
  dependence of performance on $l$ will become negligible beyond a
  point. We may view starting $r$ random walks from each node as a
  practical method to capture the steady-state behavior with small
  $l$. The most significant parameter appears to be $w$ which directly
  controls the size of the local neighborhood around each node from
  which the set of positive node-pairs are formed. Out results
  indicate the performance is poor when $w$ is too small, but plateaus
  as $w$ increases. When $w$ is extremely large, we should expect
  performance to suffer since then all node-pairs would appear in the
  positive set which would then resemble the positive set of
  node-pairs from a completely connected graph which has no community
  structure. The performance also appears to improve with increasing
  embedding dimension $d$ up to a point and then plateaus. Although
  the node embedding algorithm is not explicitly optimized for
  community discovery, embeddings that work well for community
  discovery via Euclidean-space clustering should be such that the
  embedding vectors of nodes from the same community should be roughly
  equidistant from each other. The distances of embedding vectors from
  one community to those of another community should also be roughly
  similar. These conditions are harder to meet in lower dimensions
  unless the embedding vectors from the same community are all
  identical. In higher dimensions there are more degrees of freedom
  available for the distance properties to be satisfied.}

\revision{\noindent{\bf Algorithms for performance comparison:} In the
  rest of this paper, we compare the proposed ``VEC'' algorithm
  against two baseline approaches: $(1)$ Spectral Clustering (SC) that
  is widely adopted in practice \cite{SC1large,SC2RCY,SC3LR, SC4Mc}
  and $(2)$ Acyclic Belief Propagation (ABP) which can achieve the
  information-theoretic limits in SBMs \cite{decelle2011KS, abbe2015,
    abbe2016nips}. We also include a limited comparison with another
  state-of-the-art algorithm BigClam (BC) \cite{yang2013overlapping}
  suggested by one of the reviewers.
For SC we use a state-of-the-art implementation that can handle large
scale sparse graphs \cite{SC1large}.}
In order to assure the best performance for ABP, we assume that the
ground-truth SBM model parameters are known, and adopt the parameters
suggested in \cite{abbe2016nips} which are functions of the
ground-truth parameters. In other words, we allow the competing
algorithm ABP additional advantages that are not used in our proposed
VEC algorithm. \revision{Our implementation is available at
  \url{https://github.com/cy93lin/SBM_node_embedding}.}


%
\section{Experiments with the Stochastic Block Model}
\label{sec:sbm}

In this section, we present and discuss a comprehensive set
experimental results on graphs that are synthetically generated using
a Stochastic Block Model (SBM). SBMs have been widely used for both
theoretical analysis as well as empirical validation of community
detection algorithms \cite{white1976social, SBMapp,
  fortunato2010community, NMI,
  yang2016revisiting}. 

\subsection{The Stochastic Block Model and Simulation Framework}
\label{sbm-settings}

\noindent{\bf Generative procedure}: In an SBM, a random graph with
$K$ latent communities is generated thus: $(1)$ Each node $i$ is
randomly assigned to one community $\pi_{i}\in\{1,...K\}$ with
community membership probabilities given by the probability vector
$\mathbf{p}=(p_1,\ldots, p_K)$; $(2)$ For each unordered pair of nodes
$\{i,j\}$, an edge is formed with probability $\mathbf{Q}_{n}(\pi_{i},
\pi_{j}) \in [0,1]$. Here, $\mathbf{Q}_{n}$ are the
self- and cross-community connection probabilities and are typically
assumed to vanish as $n\rightarrow \infty$ to capture the sparse
connectivity (average node degree $\ll n$) of most real-world networks
\cite{leskovec2008statistical}.


%
\noindent {\bf Weak and exact recovery}: We consider two definitions
of recovery studied in SBMs.
Let accuracy $\alpha$ be the fraction of nodes for which the estimated
communities $\hat{\mathbf{\pi}}$ agree with $\mathbf{\pi}$
\revision{(for the best permutation of node labels)}. Then,
\begin{itemize}
\item[$(1)$] {\it Weak} recovery is solvable if an algorithm can
  achieve accuracy $\alpha > \epsilon + \max_{k} p_{k}$, for some
  $\epsilon>0$, with probability $1-o_{n}(1)$
\item[$(2)$] {\it Exact} recovery is solvable if an algorithm can
  achieve accuracy $\alpha =1$ with probability $1-o_{n}(1)$.
\end{itemize}

\noindent{\bf Simulation setting and scaling regimes}: 
In the bulk of our experiments, we synthesize graphs with {\it
  balanced} communities, i.e., $\mathbf{p} = (1/K,\ldots,1/K)$, and
{\it equal} community connection probabilities. Specifically, for
$\mathbf{Q}_{n}$, we consider the standard {\it planted partition
  model} where $Q_{n}(1,1) = \ldots = Q_{n}(K,K)$ and
$Q_{n}(k,k^{\prime})$, for all $k \neq k^{\prime}$, are the same.
%
%
In Sec.~\ref{sec:unbalance}, we study how {\it unbalanced} communities
and unequal connectivities affect the performance of different
algorithms.

We consider two commonly studied scaling regimes and parameter
settings for $Q_{n}$, namely
\begin{itemize}
\item[$(i)$] {\it constant expected node degree scaling:} $Q_{n}(k,k)
  = \frac{c}{n}$, $Q_{n}(k,k^{\prime}) = \frac{c(1 - \lambda)}{n}$ and
\item[$(ii)$] {\it logarithmic expected node degree scaling:}
  $Q_{n}(k,k) = \frac{c^{\prime}\ln(n)}{n}$, $Q_{n}(k,k^{\prime}) =
  \frac{c^{\prime}\ln(n)(1-\lambda)}{n}$.
%
\end{itemize}
Intuitively, $c$ and $c^{\prime}$ influence the degree of sparsity
whereas $\lambda$ controls the degree of separation between
communities.  Let $\mu := 1+(K-1)(1-\lambda)$. The constant expected
node degree scaling regime is more challenging for community recovery
than the logarithmic expected node degree regime.
The most recent results in \cite{BP14, abbe2015, abbe2016nips} when
specialized to the {\it planted partition model} can be summarized as
follows:

\noindent{\it Condition 1}:~ For constant scaling, weak recovery is
guaranteed if \hbox{$ \frac{\lambda^{2}c}{K\mu}>1$}. For $K \leq 4$,
the condition is also necessary.
 
\noindent{\it Condition 2}:~ For logarithmic scaling, exact recovery
is solvable if, and only if, $ \sqrt{c^{\prime}}
-\sqrt{c^{\prime}(1-\lambda)} > \sqrt{K}$.

We choose different combinations of $c, c^{\prime}, K, \lambda$ in
order to explore recovery behavior around the weak and exact recovery
thresholds.
We set $\lambda = 0.9$ in both cases as it is typical in real-world
datasets ({\it cf.} Sec.~\ref{subsec:real}). For each combination of
model parameters, we synthesize $5$ random graphs and report the mean
and standard deviation of all the performance metrics (discussed
next).
%

\noindent {\bf Performance metrics}: In all our experiments, we adopt
the commonly used {\bf Normalized Mutual Information} (NMI) \cite{NMI}
and {\bf Correct Classification Rate} (CCR) metrics
\cite{murphy2012machine} to measure the clustering accuracy since
ground-truth community assignments are available.
%
%
\revision{For all $y, \hat{y} \in \{1,\ldots,K\}$, let $n_{y \hat{y}}$
  denote the number of (ground-truth) community-$y$ nodes that are
  labeled as community-$\hat{y}$ by some community discovery
  algorithm.  Then the CCR is defined by:
\[
CCR := \frac{1}{n} \sum_{k=1}^K n_{kk}.
\]
To define NMI, let $p_{Y\widehat{Y}}(y,\hat{y}) := \frac{n_{y
    \hat{y}}}{n}$ denote the empirical joint pmf of the ground-truth
and the estimated labels and $p_{Y}(y)$ and $p_{\widehat{Y}}(\hat{y})$
their marginals. Then,
\[
NMI := \frac{I(Y;\widehat{Y})}{(H(Y)+H(\widehat{Y}))/2},
\]
where
\[
I(Y;\widehat{Y}) := \sum_{y=1}^K\sum_{\hat{y} = 1}^K
p_{Y\widehat{Y}}(y,\hat{y}) \log
\left(\frac{p_{Y\widehat{Y}}(y,\hat{y})}{p_{Y}(y)p_{\widehat{Y}}(\hat{y})}\right)
\]
(with the convention $0 \log (\cdot ) = 0$) is the mutual information
between $Y$ and $\widehat{Y}$ and $H(Y)$ and $H(\widehat{Y})$ are
their entropies:
\[
H(Y) := \sum_{y=1}^{K} p_{Y}(y) \log\left(\frac{1}{p_{Y}(y)}\right),
\]
\[
H(\widehat{Y}) := \sum_{\hat{y}=1}^{K} p_{\widehat{Y}}(\hat{y})
\log\left(\frac{1}{p_{\widehat{Y}}(\hat{y})}\right).
\]
Both CCR and NMI are symmetric with respect to the ground-truth labels
$Y $and the estimated labels $\widehat{Y}$. However {\bf NMI is
  invariant to any permutation of labels, but CCR is not}. We
therefore calculate CCR based on the best re-labeling of the estimated
labels, i.e.,
\[
\max_{\sigma = \text{label permutation}} \frac{1}{n} \sum_{y=1}^K n_{y\sigma(y)}.
\]
}
In order to compare the {\it overall relative performance} of
different algorithms across a large number of different simulation
settings, we also compute the NMI and CCR {\bf Performance Profile}
(PP) curves \cite{PP} across a set of 250 distinct experiments. These
curves provide a global performance summary of the compared
algorithms.

Table~\ref{exptlandscape} provides a bird's-eye view of all our
experiments with synthetically generated graphs. The \revision{table
  presents} all key problem parameters that are held fixed as well as
those which are varied. It also summarizes the main conclusion of each
experimental study and includes pointers to the appropriate figures
and subsections where the results can be found.
%
\begin{table*}[!htb]
\caption{Summary of all experiments with synthetic graphs \label{exptlandscape}}
\centering
\renewcommand{\arraystretch}{1.5}
\begin{tabular}{|c|c|c|c|c|p{0.111\linewidth}|c|p{0.24\linewidth}|}
\hline 
\# & Fig., Table \& Sec.& Scaling regime & Sparsity  $c$ or $c^{\prime}$ & Graph size $n$ &
Balanced ${\bf p}$ \& unifrm. $\text{diag}(\mathbf{Q})$? & $K$ & Main observation \\ 
\hline \hline
1 & Fig.~\ref{fig:transition_1}, Sec.~\ref{weak-recovery} & constant & variable & $1e4$ & yes & $2$ &  VEC exhibits weak recovery phase transition behavior \\ 
\hline 
2& Fig.~\ref{fig:transition_1_n}, Sec.~\ref{weak-recovery} & constant & fixed & variable & yes & $2$ & VEC achieves weak recovery asymptotically when conditions are satisfied\\ 
\hline 
3& Fig.~\ref{fig:cross_weak_k5}, Sec.~\ref{weak-recovery} & constant & variable & $1e3$ & yes & $5$ & VEC can cross the weak recovery limit for $K > 4$ \\ 
\hline
4& Fig.~\ref{fig:Kvar}, Sec.~\ref{weak-recovery} & constant & fixed & $1e4$ & yes & variable & VEC is robust to the number of communities $K$\\ 
\hline 
5& Fig.~\ref{fig:gamma-var}, Sec.~\ref{sec:unbalance} & constant & fixed & $1e4$ & unbalanced $\mathbf{p}$ & $2$ & VEC is robust to unbalanced communities \\ 
\hline 
6& Fig.~\ref{fig:beta-var}, Sec.~\ref{sec:unbalance} & constant & fixed & $1e4$ & non-uniform
$\text{diag}(\mathbf{Q})$ & $2$ & VEC is robust to unequal connectivities\\ 
\hline 
7& Fig.~\ref{fig:strong_trans}, Sec.~\ref{exactrecovery} & logarithmic & variable & $1e4$ & yes  & $2$ & VEC attains the exact recovery limit \\ 
\hline 
8& Fig.~\ref{fig:logarithm_scaling}, Sec.~\ref{exactrecovery} & logarithmic & fixed & variable & yes & $2$ & VEC achieves exact recovery asymptotically when conditions are satisfied\\ 
\hline 
9& Fig.~\ref{fig:params}, Sec.~\ref{sec:param-sensitive} & logarithmic & fixed & $1e4$ & yes & $5$ & VEC is robust to algorithm parameters \\ 
\hline 
10& Table~\ref{table:random}, Sec.~\ref{sec:param-sensitive} & both &
fixed & $1e4$ & yes & 2, 5 & VEC is robust to randomness in creating paths \\ 
\hline 
11& Fig.~\ref{fig:PP}, Sec.~\ref{performanceprofiles} & constant & variable & variable & yes & variable & VEC consistently outperforms baselines across 240 experiments \\ 
\hline
12& Table~\ref{table:dc-sbm}, Sec.~\ref{dc-result} & both & variable & $1e3$ & yes & variable & VEC outperforms baselines on degree-corrected model \\ 
\hline
\end{tabular} 
\end{table*}
%


\subsection{Weak Recovery Phase Transition}
\label{weak-recovery}
\begin{figure}[!htb]
\centering
\includegraphics[width=0.75\linewidth]{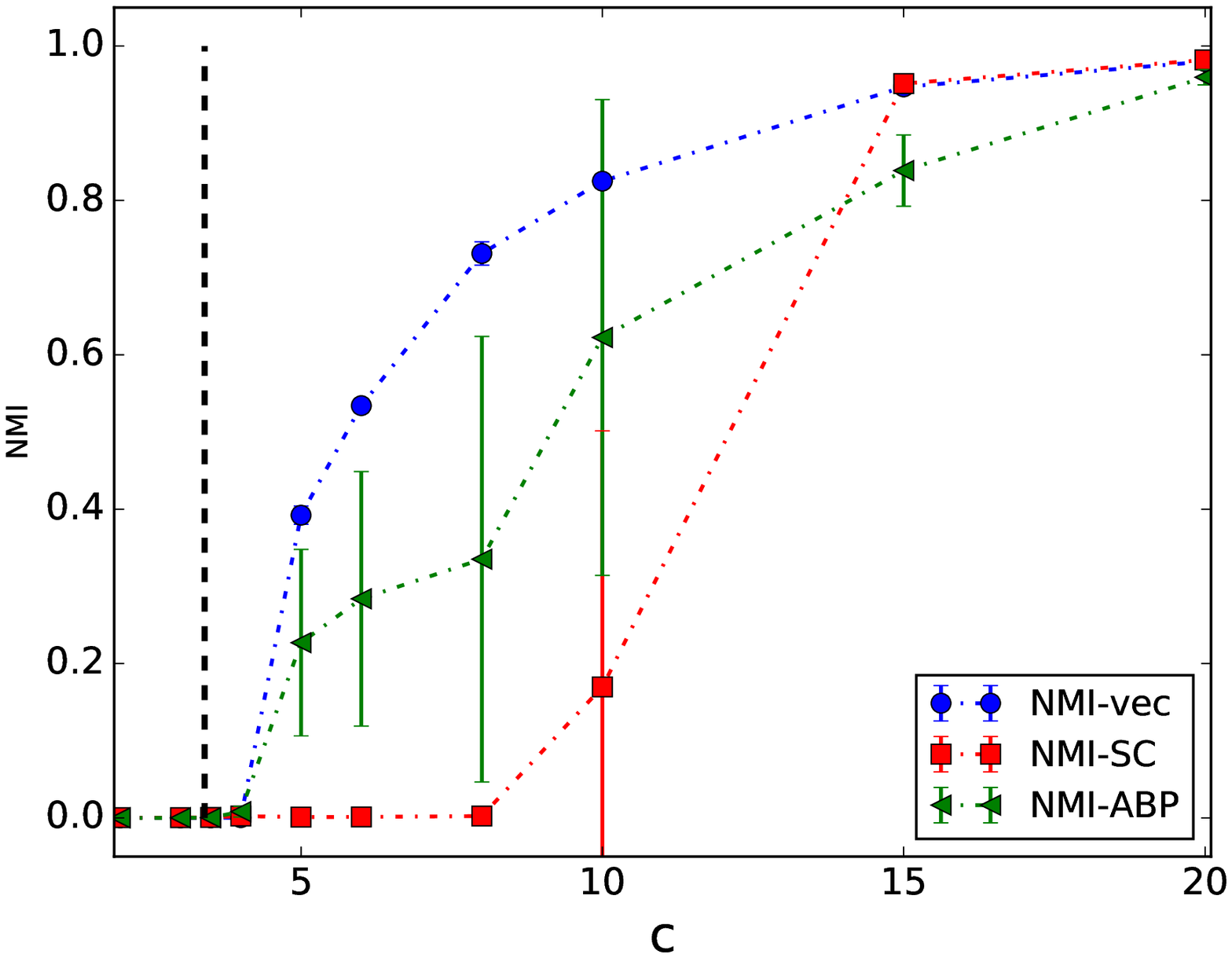}
\includegraphics[width=0.75\linewidth]{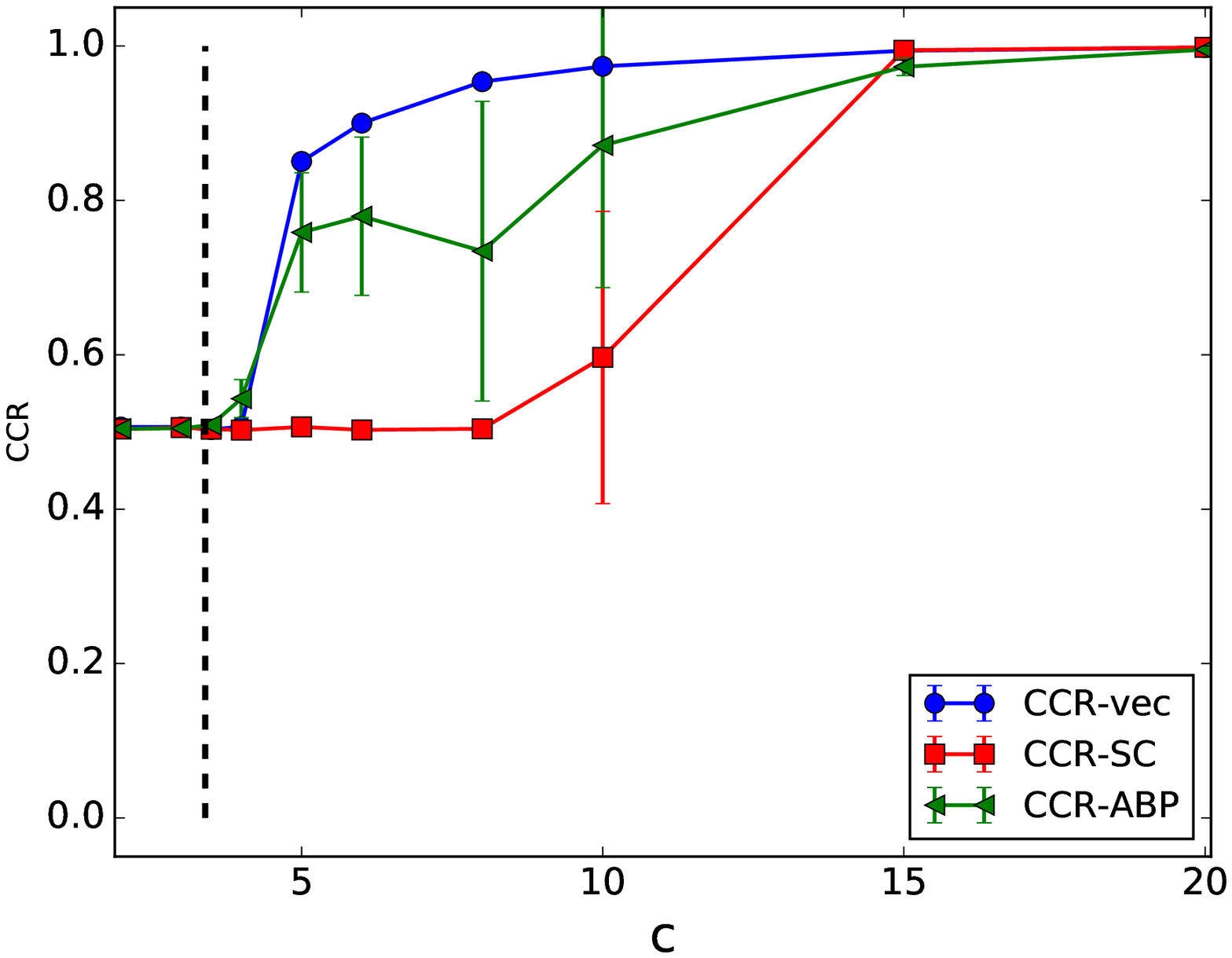} 	
\caption{\small NMI (\revision{top sub-figure}, dashed curves) and CCR
  (\revision{bottom sub-figure}, solid curves) versus sparsity level
    $c$ for VEC, SC, and ABP on SBM graphs with constant degree
    scaling. Here, $\mathbf{p}$ is uniform, $K=2, \lambda = 0.9$, and
    $n=10000$. The vertical dashed line is $c_{\text{weak}}=
    2.8$. This figure is best viewed in color.}
\label{fig:transition_1}
\end{figure}
To understand behavior around the weak recovery limit, we synthesized
SBM graphs with $K=2$, $n=10000$, and $\lambda = 0.9$ at various
sparsity levels $c$ in the constant scaling regime.
For these parameter settings, weak recovery is possible if, and only
if, $c > c_{\text{weak}} \approx 2.8$ ({\it cf. Condition 1}). The
results are summarized in Fig.~\ref{fig:transition_1}.

Figure~\ref{fig:transition_1} reveals that the proposed VEC algorithm
exhibits {\it weak recovery phase transition} behavior: for
$c>c_{\text{weak}}$, $\text{CCR} > 0.5$ and when $c< c_{\text{weak}}$,
$\text{CCR} \approx 0.5$ (random guess).
This behavior can be also observed through the NMI metric. 
%
The behavior of ABP which provably achieves the weak recovery limit
\cite{abbe2016nips} is also shown in
Fig.~\ref{fig:transition_1}. Compared to ABP, VEC has consistently
superior mean clustering accuracy over the entire range of $c$ values.
In addition, we note that the variance of NMI and CCR for ABP is
significantly larger than VEC.
This is discussed later in this section.
SC, however, does not achieve weak recovery for sparse $c$ ({\it
  cf.}~Fig.~\ref{fig:transition_1}) which is consistent with theory
\cite{SC3LR}.
\begin{figure}[!htb]
\centering
\includegraphics[width=0.75\linewidth]{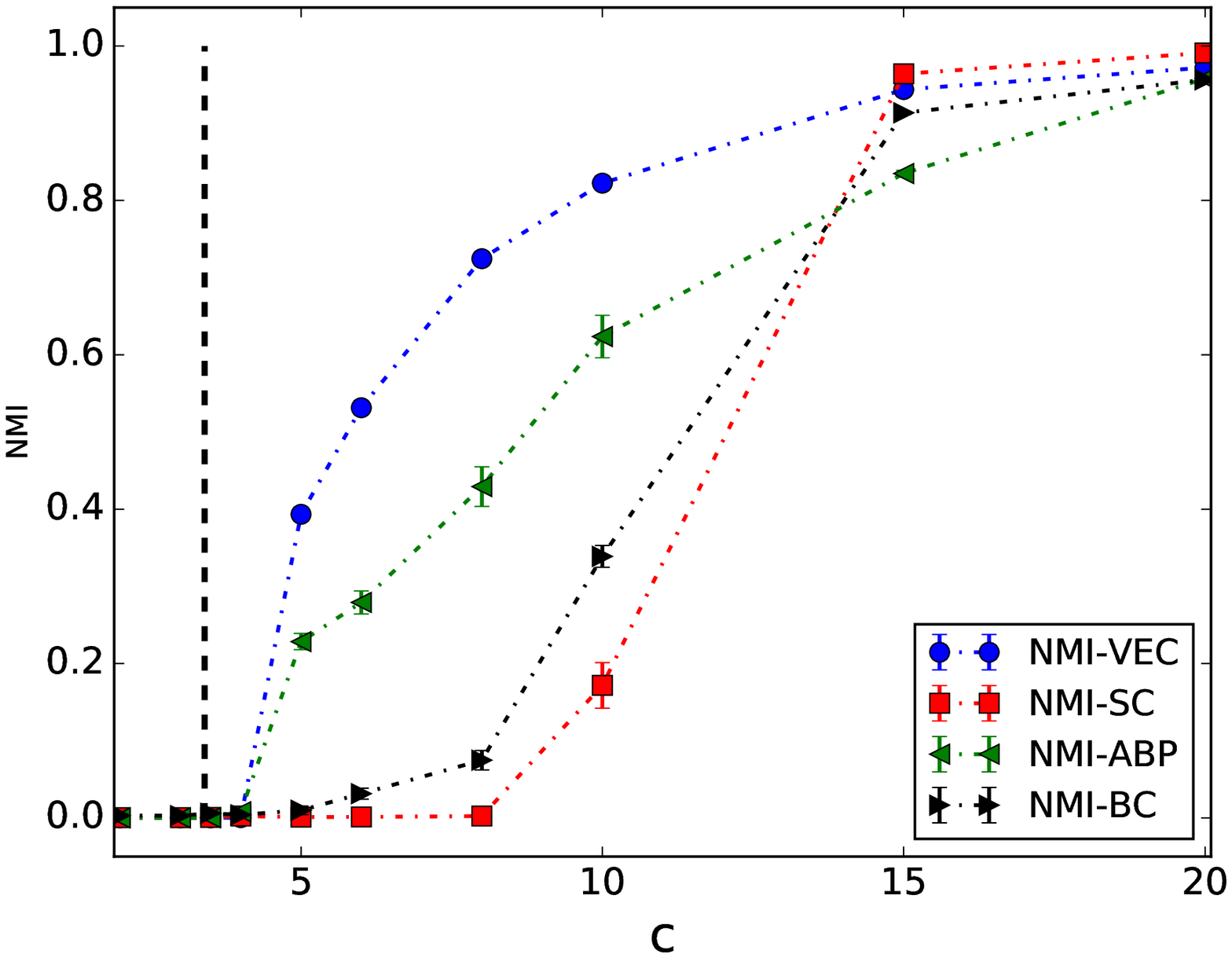}
\includegraphics[width=0.75\linewidth]{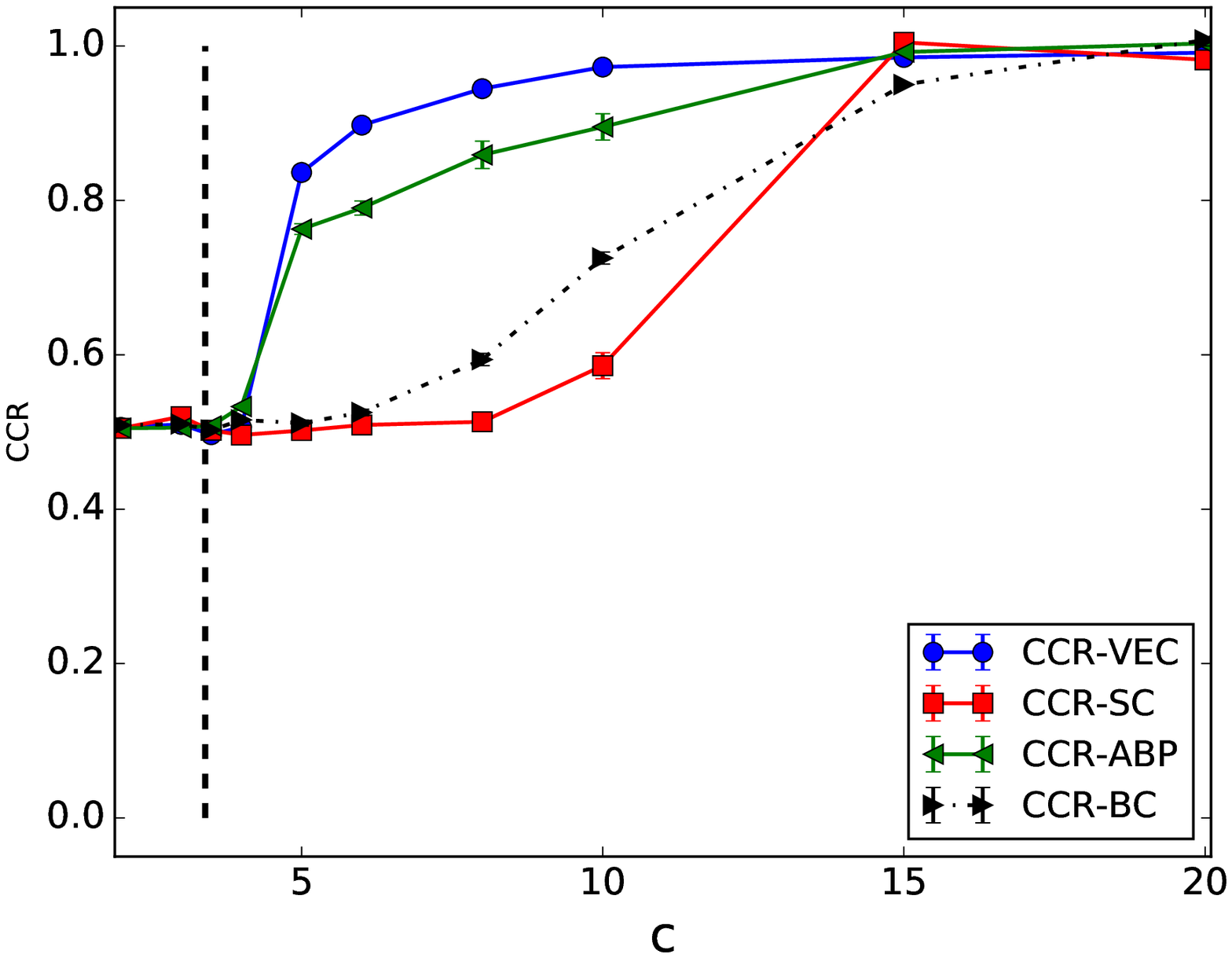} 	
\caption{\revision{\small NMI (\revision{top sub-figure}, dashed
    curves) and CCR (\revision{bottom sub-figure}, solid curves)
    versus sparsity level $c$ for VEC, SC, and ABP on SBM graphs with
    constant degree scaling. Here, $\mathbf{p}$ is uniform, $K=2,
    \lambda = 0.9$, and $n=10000$. The vertical dashed line is
    $c_{\text{weak}}= 2.8$. Each CCR/NMI value is based on $500$
    independent random graph realizations. Each point represents the
    mean CCR/NMI value and the associated confidence bar
    ($({\text{empirical std\_dev}})/\sqrt{500}$) are shown. The
    performance of BigClam (BC, \cite{yang2013overlapping}) is also
    shown. This figure is best viewed in color.} }
\label{fig:transition_1_500}
\end{figure}
\revision{\noindent{\bf Effect of increasing number of random
    experiments}:
The curves shown in Fig.~\ref{fig:transition_1} are based on averaging
performance metrics across multiple, independently generated, random
realizations of SBM graphs. In order to understand the impact of the
number of realizations on the overall performance trends of different
algorithms, we increased the number of random graphs used to create
each data point in Fig.~\ref{fig:transition_1} from $5$ to $500$. The
resulting mean NMI and CCR values and their associated confidence
intervals are shown in Fig.~\ref{fig:transition_1_500}.  Comparing the
curves in Fig.~\ref{fig:transition_1} and
Fig.~\ref{fig:transition_1_500}, we see that they are very similar.
The confidence intervals of all algorithms have clearly diminished as
expected. Thery were already very small for VEC even with $5$
realizations. They have all but ``disappeared'' for VEC with $500$
realizations. The only significant change is to the mean curve for ABP
which has become monotonic (as it should) and smoother after
increasing the number of random realizations. Since the curves with
$5$ and $500$ realizations are so similar, we only use $5$ random
realizations for each SBM parameter setting in the rest of this paper.}

\revision{\noindent{\bf Comparison with
    BigClam~\cite{yang2013overlapping}}:
Figure~\ref{fig:transition_1_500} also shows the performance of
BigClam \cite{yang2013overlapping} (BC) a recent powerful community
detection algorithm based on matrix factorization techniques that
scales well to large graphs of millions of nodes and edges.  We
observe that BC outperforms SC when the graph is relatively
sparse. But VEC and ABP still outperform BC. Since the performance of
BigClam is quite similar to that of SC, we decided to exclude BC in
the remainder of our experiments.}

\begin{figure}[htb!]
\centering
\includegraphics[width=0.75\linewidth]{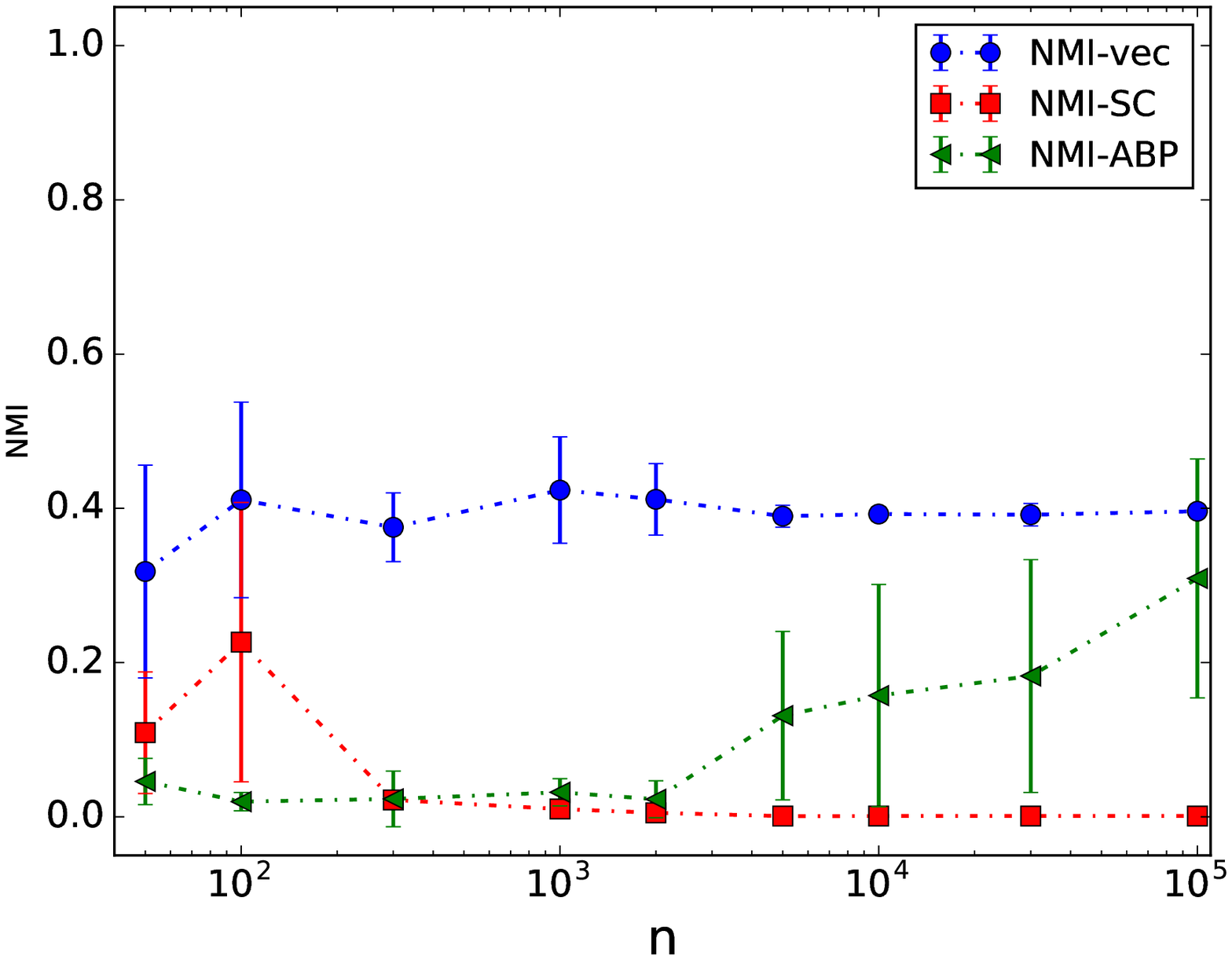} 
\includegraphics[width=0.75\linewidth]{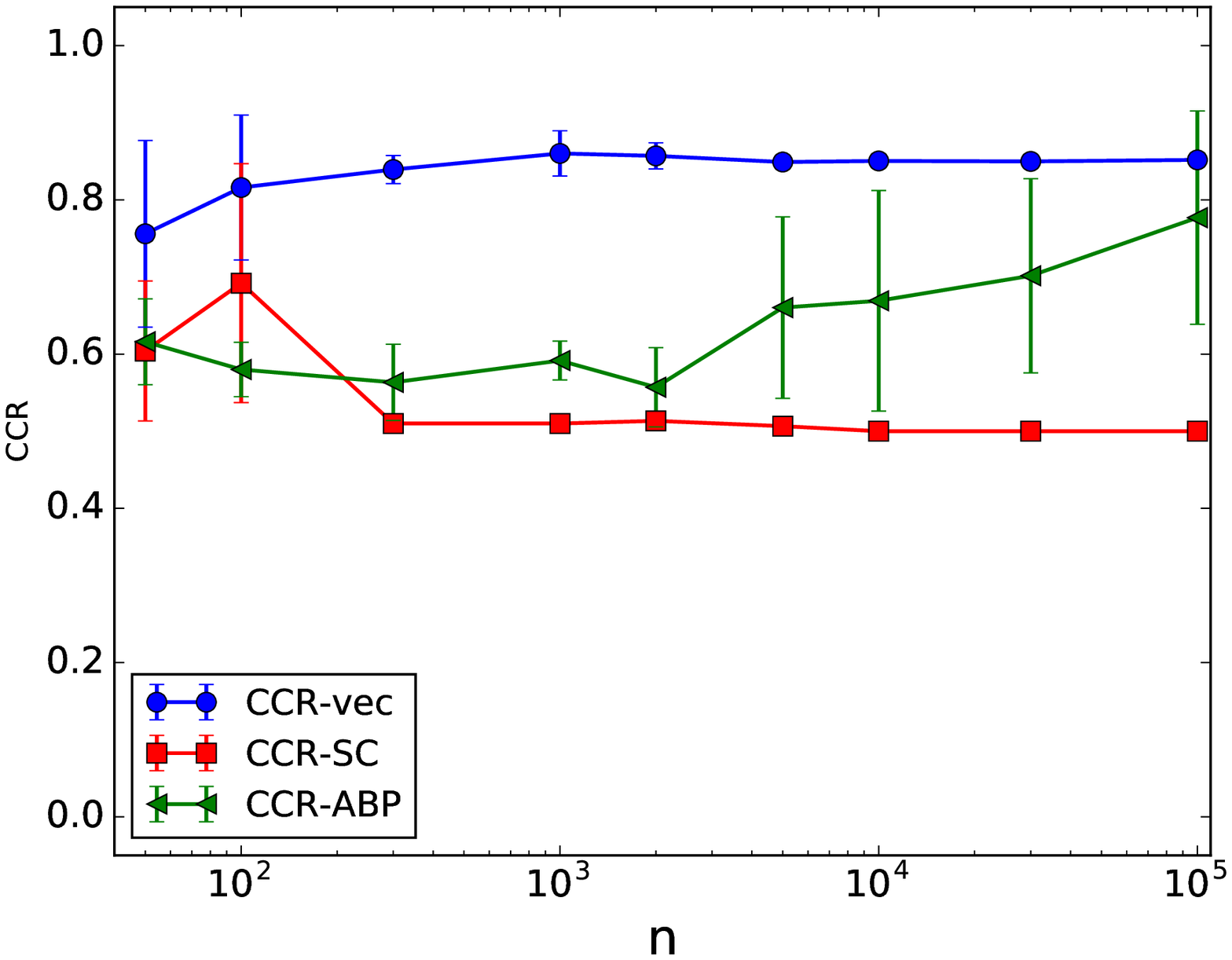} 	
\caption{\small NMI (\revision{top sub-figure}, dashed curves) and CCR
  (\revision{bottom sub-figure}, solid curves) versus $n$ for VEC, SC,
  and ABP on SBM graphs with constant degree scaling. Here,
  $\mathbf{p}$ is uniform, $K=2, \lambda = 0.9$, and $c=5.0 >
  c_{\text{week}}$.}
\label{fig:transition_1_n}
\end{figure}
\revision{\noindent{\bf Effect of graph size:}} In order to
\revision{provide further support for the observations presented
  above} we also synthesized SBM graphs with increasing graph size $n$
with $K=2, \lambda=0.9, c= 5.0$ held fixed in the constant degree
scaling regime. Since $c= 5.0 >c_{\text{weak}}$, weak recovery is
possible asymptotically as $n \rightarrow \infty$.
As shown in Fig.~\ref{fig:transition_1_n}, VEC can empirically achieve
weak recovery for both small and large graphs, and consistently
outperforms ABP and SC. While ABP can provably achieve weak recovery
asymptotically, its performance on smaller graphs is poor. \\

%
\noindent {\bf Crossing below the weak recovery limit for $K>4$}:
\begin{figure}[htb!]
\centering
\includegraphics[width=0.75\linewidth]{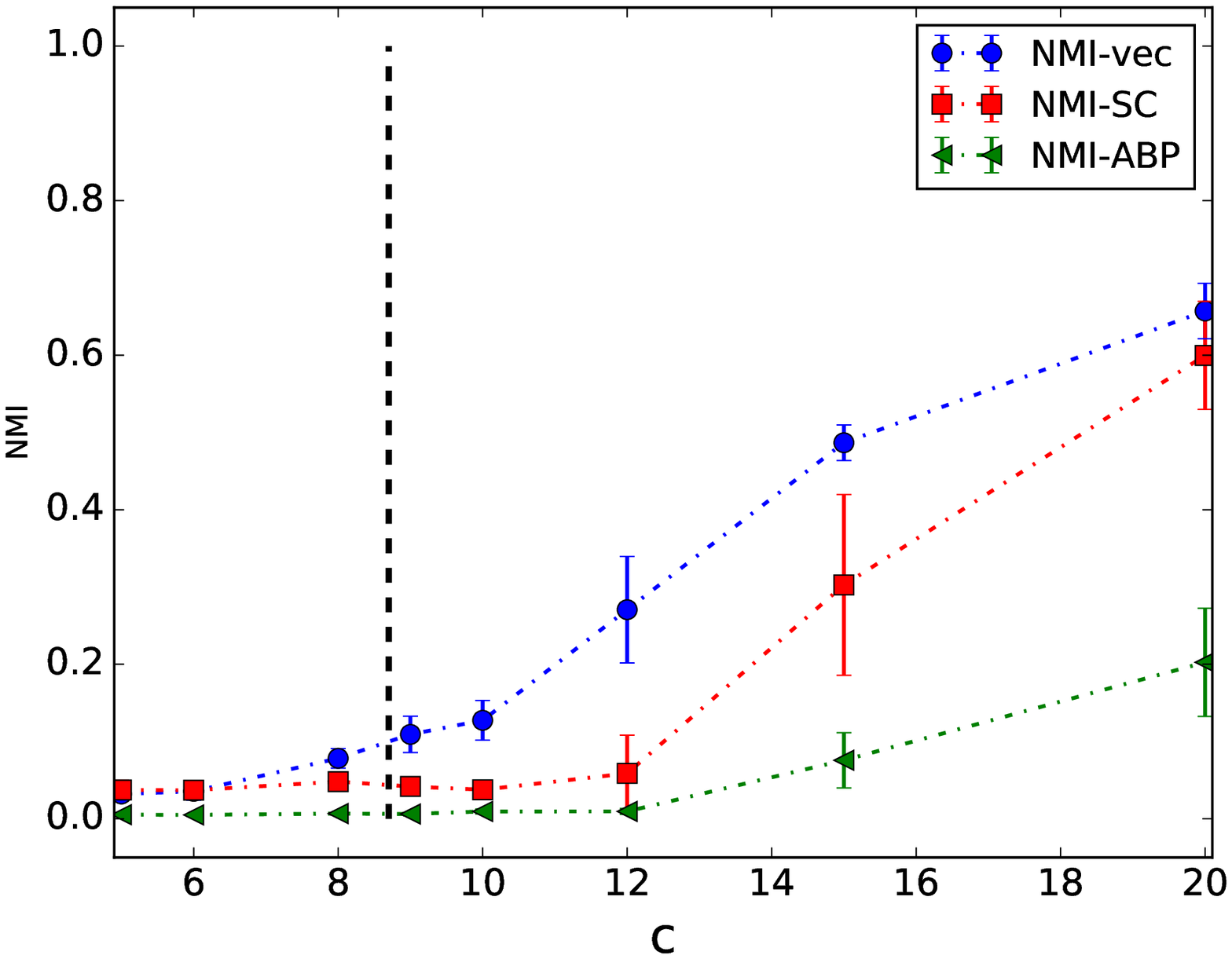}
\includegraphics[width=0.75\linewidth]{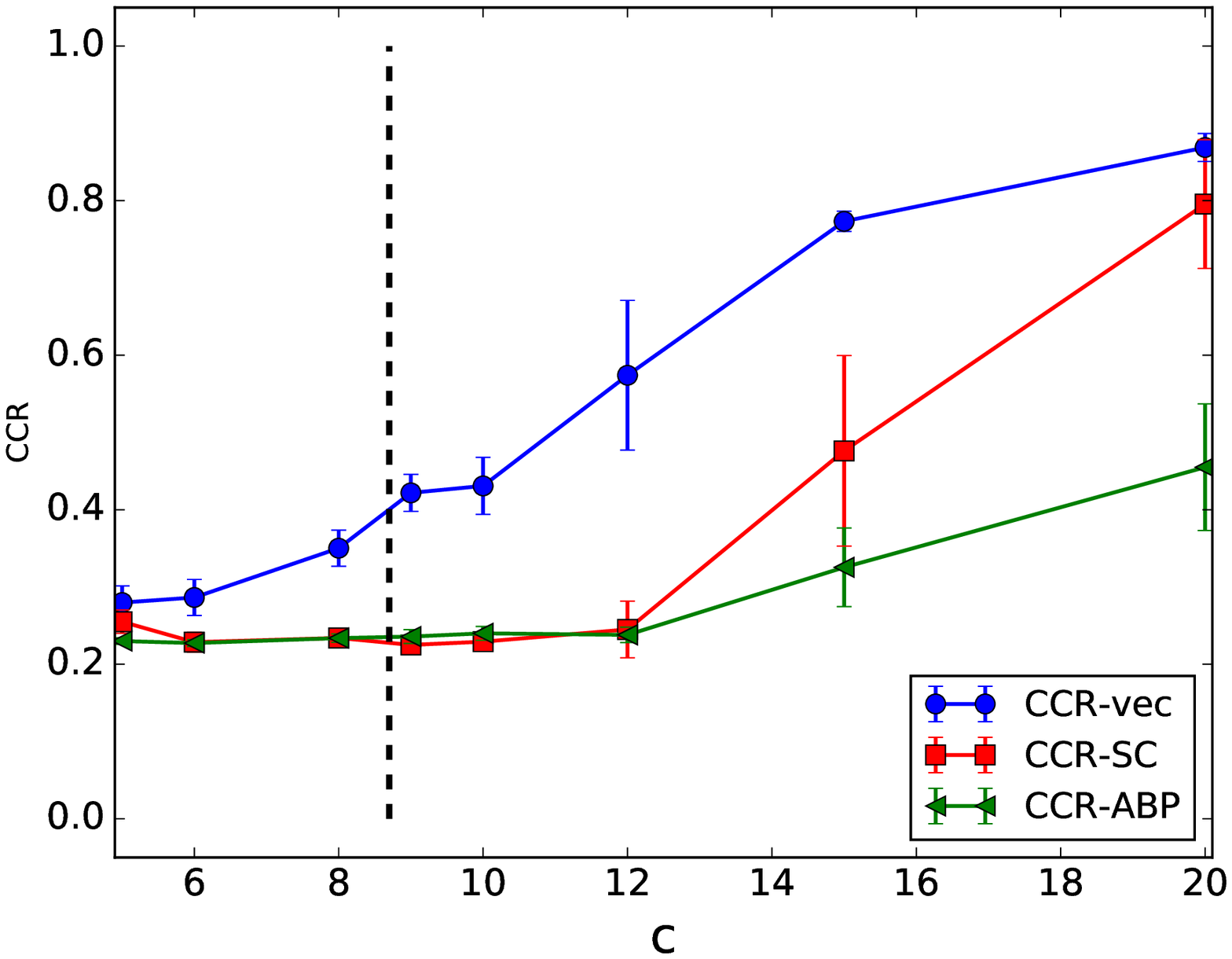} 	
\caption{\small NMI (\revision{top sub-figure}, dashed curves) and CCR
  (\revision{bottom sub-figure}, solid curves) versus sparsity level
  $c$ for VEC, SC, and ABP on SBM graphs with constant degree
  scaling. Here, $\mathbf{p}$ is uniform, $K=5, \lambda = 0.9$, and
  $n=1000$. The vertical dashed line is $c_{\text{weak}} = 8.6$.}
\label{fig:cross_weak_k5}
\end{figure}

Here we explore the behavior of VEC below the weak recovery limit for
$K>4$ since to-date there are no necessary and sufficient weak
recovery bounds established for this setting (i.e., $K > 4$).
Similar to Fig.~\ref{fig:transition_1}, we synthesized SBM graphs in
the constant degree scaling regime for various sparsity levels $c$
fixing $K=5, \lambda =0.9$, and $n=1000$. In this setting, $c >
c_{\text{weak}} = 8.7$ is {\it sufficient but not necessary} for weak
recovery. The results are summarized in Fig.~\ref{fig:cross_weak_k5}.

As can be seen in Fig.~\ref{fig:cross_weak_k5}, the VEC algorithm can
cross the weak recovery limit: for some $c\leq c_{\text{weak}}$,
$\text{CCR} > \frac{1}{K}$ and $\text{NMI} > 0$ with a significant
margin.
Here too we observe that VEC consistently outperforms ABP and SC with
a large margin. \\

\noindent{\bf Weak recovery with increasing number of communities $K$}:
Next we consider the performance of VEC as the number of communities
$K$ increases. In particular, we synthesize {\it planted partition
  model} SBMs in the constant scaling regime with $c=10$, $\lambda =
0.9$, $N=10000$, and uniform $\mathbf{p}$.
%
\revision{As $K$ increases, recovery becomes impossible, because}
according to {\it Condition 1} for weak recovery \cite{abbe2016nips},
weak recovery is possible if
$\lambda^2 c > K(1+(K-1)(1-\lambda))$. 
For the above parameter settings, we have $K\leq K_{\text{weak}} = 5$.
\begin{figure}[htb!]
\centering
\includegraphics[width=0.75\linewidth]{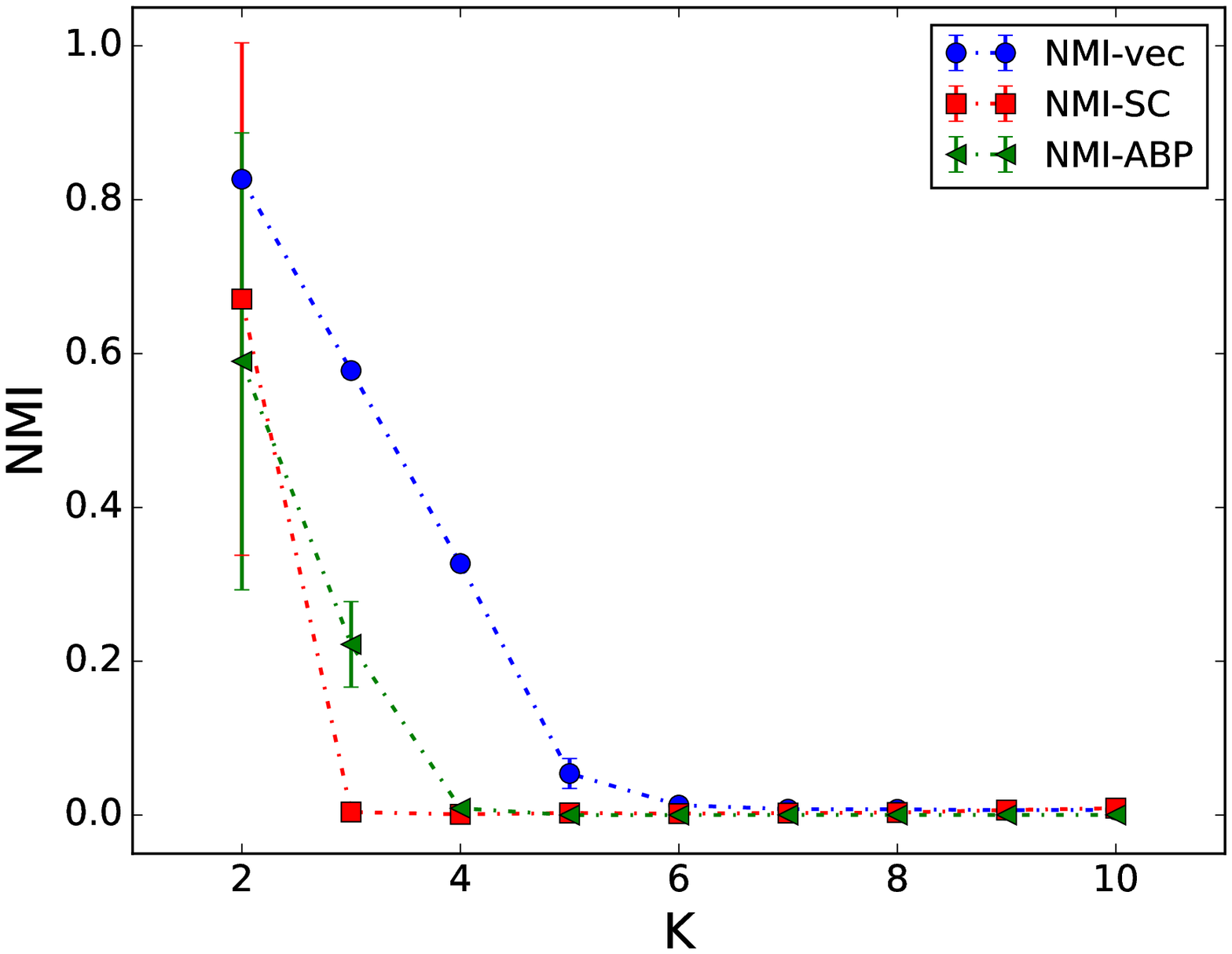}
\includegraphics[width=0.75\linewidth]{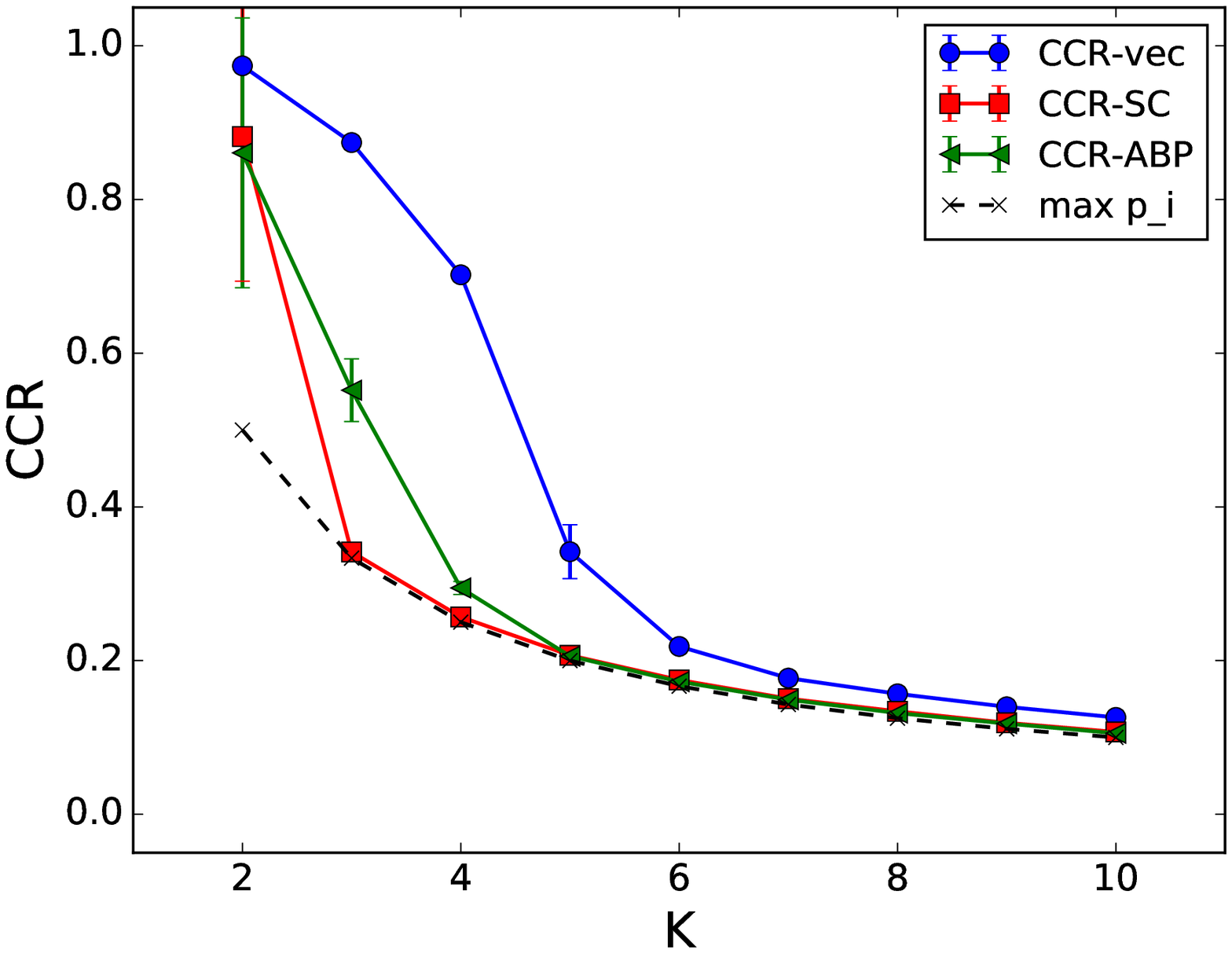}
%
\caption{\small NMI (\revision{top sub-figure}, dashed curves) and CCR
  (\revision{bottom sub-figure}, solid curves)
 versus number of communities $K$ for VEC, SC, and ABP on SBM graphs
 with constant degree scaling. Here, $\mathbf{p}$ is uniform, $\lambda
 = 0.9$, $c=10$, and $n=10000$. Weak recovery is possible if $K \leq
 K_{\text{weak}} = 5$. The black dashed curve in the CCR sub-figure is
 the plot of the maximum community weight $\max_k p_k$ versus $K$. It
 is the CCR of the rule which assigns the {\it apriori} most likely
 community to {\it all} nodes.}
\label{fig:Kvar}
\end{figure}
Figure~\ref{fig:Kvar} summarizes the performance of VEC, SC, and ABP
as a function of $K$. The performance of the three algorithms can be
compared more straightforwardly by focusing on the NMI metric (plots
in the upper sub-figure of Fig.~\ref{fig:Kvar}).  Similar to all the
previous studies of this section, the proposed VEC algorithm can
empirically achieve weak-recovery whenever the information-theoretic
sufficient conditions are satisfied, i.e., $\text{NMI} > 0$ with a
significant margin for all $K\leq K_{\text{weak}} = 5$.
We note that in terms of the $\text{CCR}$ metric, a ``weak'' recovery
corresponds to $\text{CCR} > \max_k p_k \approx 1/K$ since $\max_k
p_k$ it is the CCR of the rule which assigns the {\it apriori} most
likely community label to all nodes. This is empirically attained by
the VEC algorithm as illustrated in plots in the bottom sub-figure of
Fig.~\ref{fig:Kvar}.

Note that the performance of SC drops significantly beyond $K>2$. We
also note that the CCR performance margin between ABP, which is a
provably asymptotically consistent algorithm, and the best constant
guess rule ($\text{CCR} = 1/K$) is much smaller than for VEC.

\subsection{Weak Recovery with Unbalanced Communities and Unequal Connectivities}
\label{sec:unbalance}

Here we study how unbalanced communities and unequal connectivities
affect the performance of the proposed VEC algorithm.  
By unbalanced communities we mean an SBM in which the community
membership weights $\mathbf{p}$ are not uniform.  In this scenario,
some clusters will be more dominant than the others making it
challenging to detect 
\revision{small} clusters.
By unequal connectivities we mean an SBM in which $\mathbf{Q}_n(k,k)$
is not the same for all $k$ or $\mathbf{Q}_n(k,k')$ is not the same
for all $k \neq k'$. Since it is unwieldy to explore all types of
unequal connectivities, our study only focuses on unequal {\it
  self}-connectivities, i.e., $\mathbf{Q}_n(k,k)$ is not the same for
all $k$. In this scenario, the densities of different communities will
be different making it challenging to detect the sparser
communities. 
Here we compare NMI and CCR curves only for VEC and ABP but not
SC. When communities are unbalanced or the self-connectivities are
unequal we observed that SC takes an inordinate amount of time to
terminate. We decided therefore to omit NMI and CCR plots for SC from
the experimental results in this subsection.

\noindent{\bf Unbalanced communities}:
We first show results on SBMs with nonuniform community weights
$\mathbf{p}$. For simplicity, we consider SBMs with $K=2$ communities
and set $p_1 = \gamma\in \left[ 0.5, 1\right)$. Then, $p_2 =
  1-\gamma$. For the other parameters we set $c=8, N=10000, \lambda =
  0.9$.
From the general weak recovery conditions for nonuniform ${\bf p}$ in
\cite{abbe2016nips},\footnote{{\it Condition 1} in
  Sec.~\ref{sbm-settings} assumes uniform ${\bf p}$.} it can be shown
that as $\gamma \rightarrow 1$, the threshold for guaranteed
weak-recovery will be broken. Specifically, for the above parameter
settings, it can can be shown that $\gamma$ must not exceed
$\gamma_{\text{weak}} \approx 0.65$ for weak recovery.

\begin{figure}[htb!]
\centering
\includegraphics[width=0.75\linewidth]{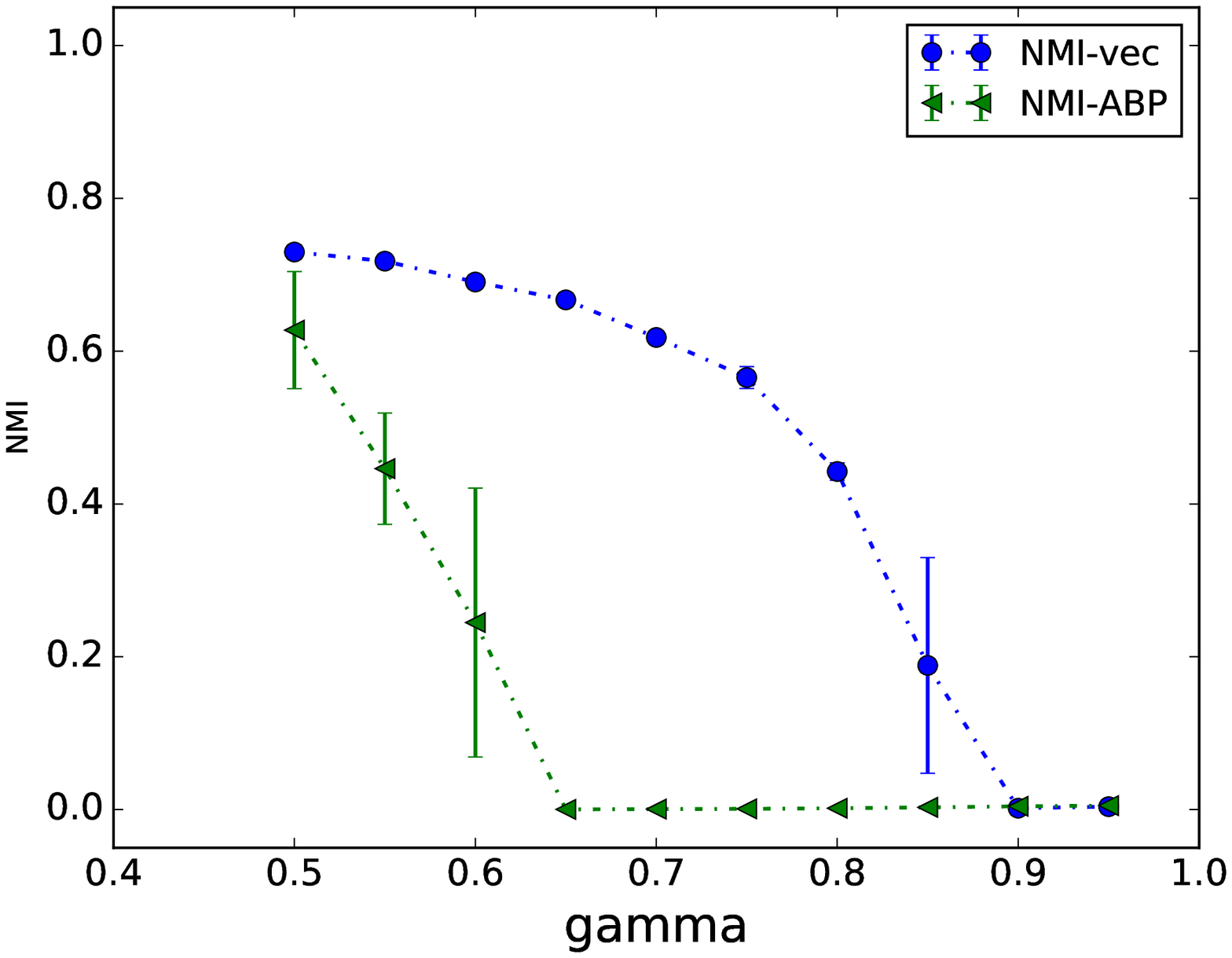}
\includegraphics[width=0.75\linewidth]{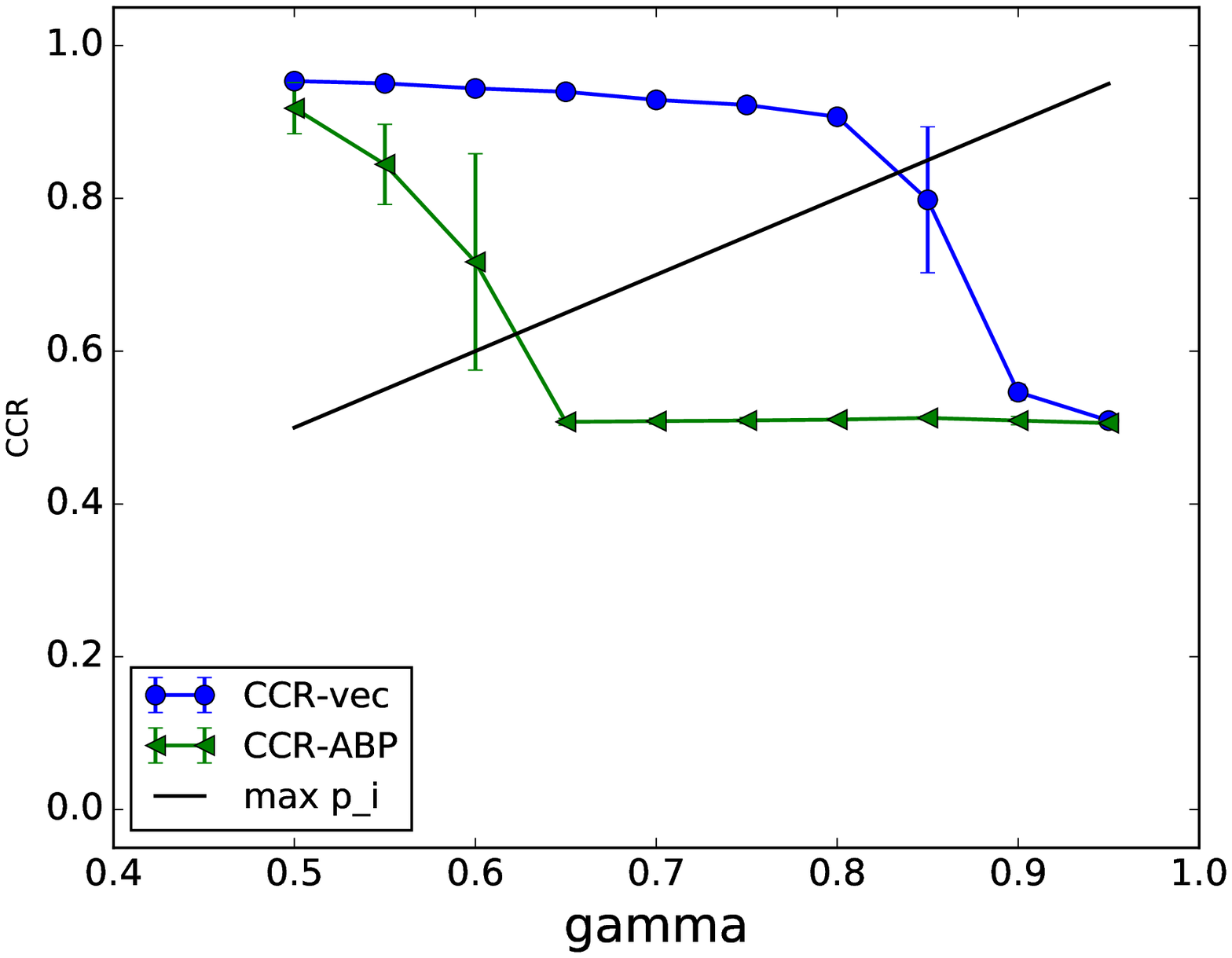}
%
\caption{\small NMI (\revision{top sub-figure}, dashed curves) and CCR
  (\revision{bottom sub-figure}, solid curves) versus maximum
  community weight $\gamma$ for VEC and ABP on SBM graphs with
  constant degree scaling. Here, $K=2$, $\mathbf{p} = \left[\gamma,
    1-\gamma \right]$, $\lambda = 0.9$, $c=8$, and $n=10000$. Weak
  recovery is possible if $\gamma \leq \gamma_{\text{weak}} =
  0.65$. The solid unmarked curve in the CCR sub-figure indicates the
  maximum community weights $\max_{k} p_k = \gamma$ in each setting.
  It is the CCR of the rule which assigns the {\it apriori} most
  likely community to {\it all} nodes.}
\label{fig:gamma-var}
\end{figure}

We summarize the results in Fig.~\ref{fig:gamma-var}.  For comparison,
the right (CCR) sub-figure of Fig.~\ref{fig:gamma-var} also shows the
plot of $\max_k p_k = \gamma$ which is the CCR of the rule which
assigns the {\it apriori} most likely community to {\it all} nodes.
%
From the figure it is evident that unlike ABP, the CCR performance of VEC
remains stable across a wide range of $\gamma$ values indicating that
it can tolerate significantly unbalanced communities.

\begin{figure}[!htb]
\centering
\includegraphics[width=0.75\linewidth]{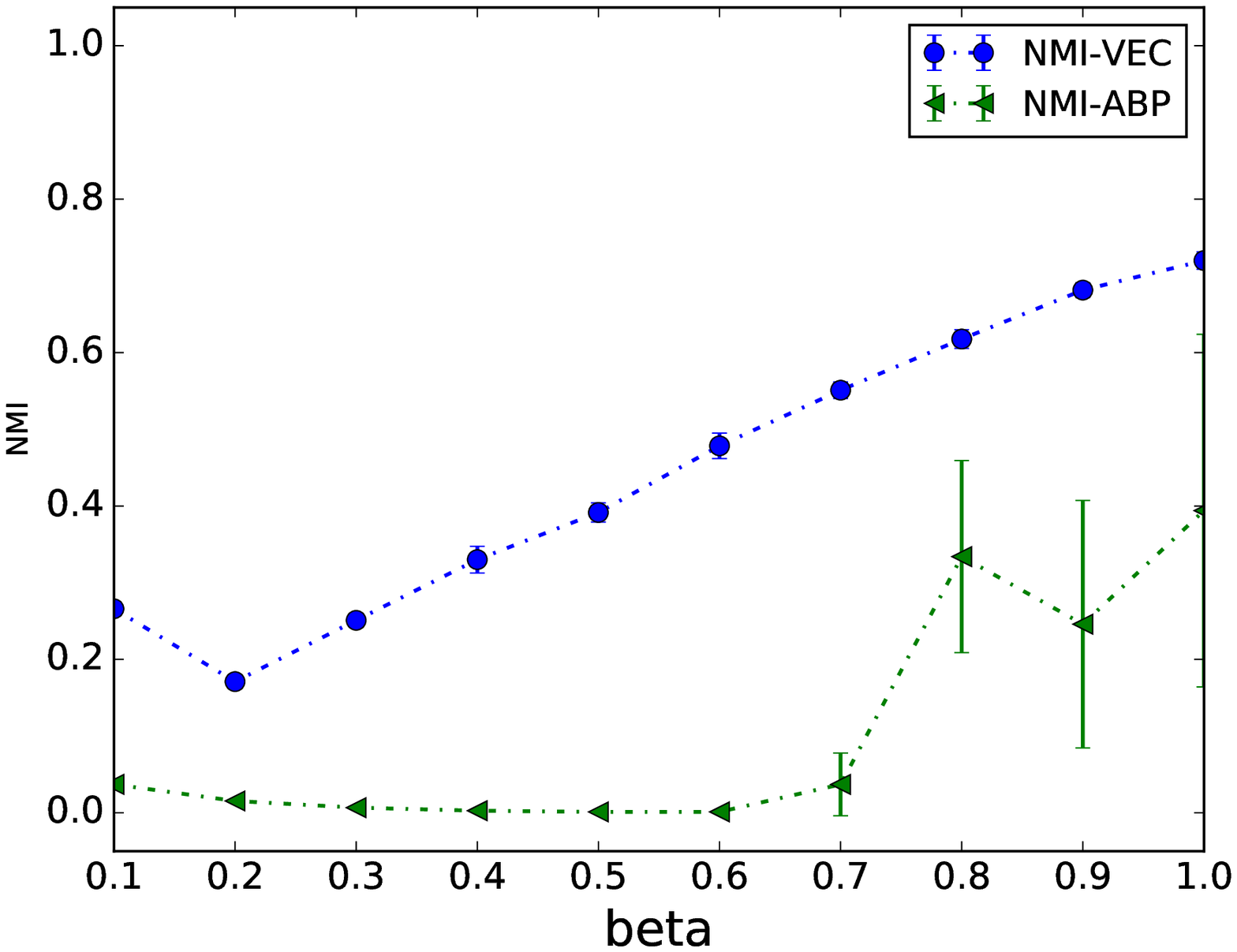}
\includegraphics[width=0.75\linewidth]{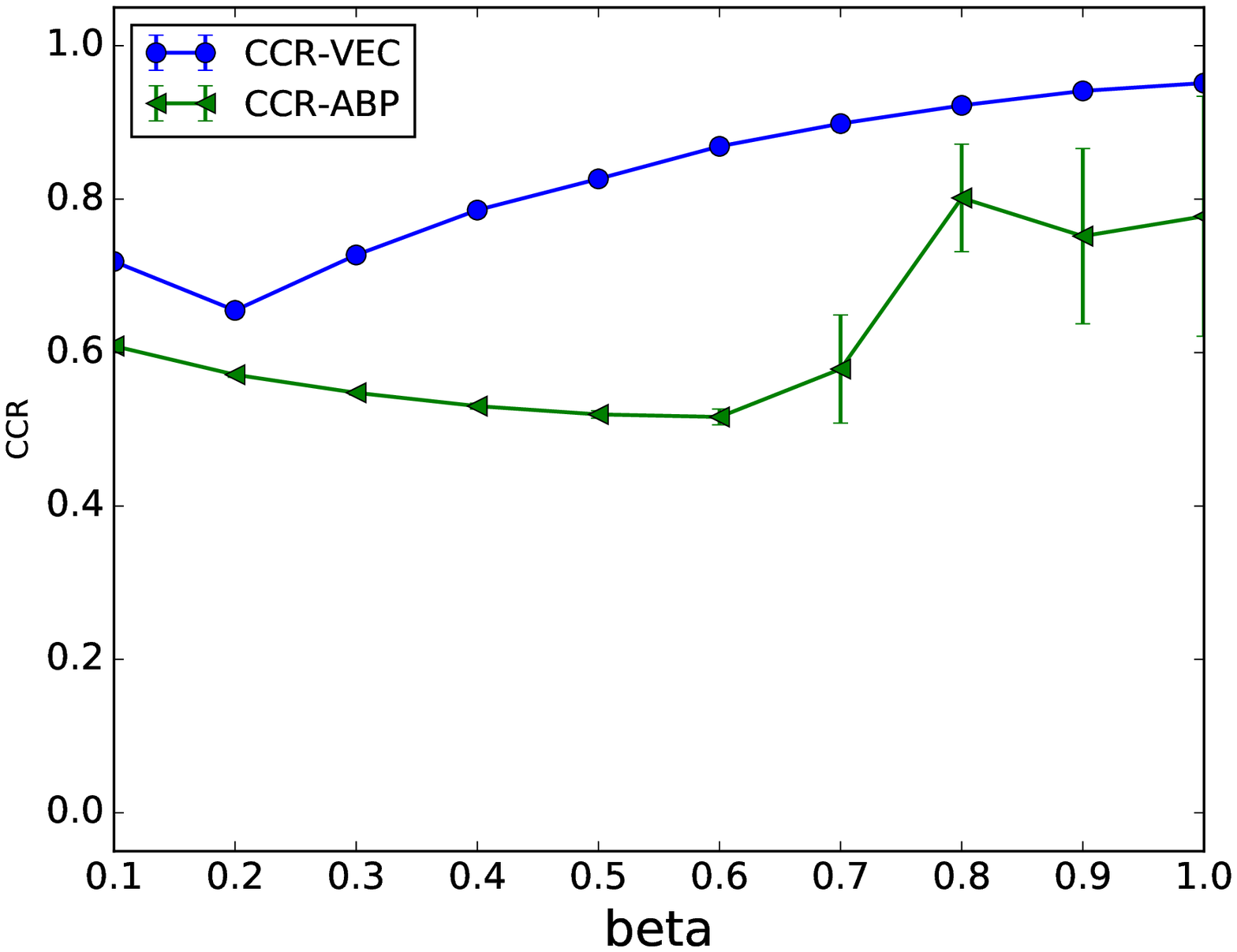}
%
\caption{\small NMI (\revision{top sub-figure}, dashed curves) and CCR
  (\revision{bottom sub-figure}, solid curves) versus community
    connectivity $\beta$ for VEC and ABP on SBM graphs with constant
    degree scaling. Here, $\mathbf{p}$ is uniform, $\lambda = 0.9$,
    $c=8$ and $n=10000$. Weak recovery is possible if $\beta \leq
    \beta_{\text{weak}} = 0.60$. The dashed curve in the CCR
    sub-figure is the maximum community weight $\max_k p_k$ in each
    setting. It is the CCR of the rule which assigns the {\it apriori}
    most likely community to {\it all} nodes.}
%
\label{fig:beta-var}
\end{figure}

\noindent{\bf Unequal community connectivity}:
We next consider the situation in which the connectivity constants of
different communities are distinct. For simplicity, we consider SBMs
with $K=2$ communities and balanced weights $p_1 = p_2 =0.5$. We focus
on the 
constant scaling regime and set
\begin{equation*}
Q_{n} = 
\begin{bmatrix}
    \frac{c}{n} & \frac{c (1-\lambda)}{n} \\
	\frac{c (1-\lambda)}{n} & \frac{c\beta}{n}
\end{bmatrix}
\end{equation*}
Here, $\beta\in\left(0, 1\right]$ determines the relative densities of
communities $1$ and $2$. For the other model parameters, we set $c=8$,
$\lambda = 0.9$ as in the previous subsection.
From the general weak recovery conditions for unequal community
connectivity in \cite{abbe2016nips}, it can be shown that 
weak recovery requires that $\beta \geq \beta_{\text{weak}} \approx
0.60$. We summarize the results in Fig.~\ref{fig:beta-var}.
From the figure it is once again evident that the performance of VEC
remains stable across a wide range of $\beta$ values when compared to
ABP.

%
%
%
%

\subsection{Exact Recovery Limits}
\label{exactrecovery}
\begin{figure}[htb!]
\centering
\includegraphics[width=0.75\linewidth]{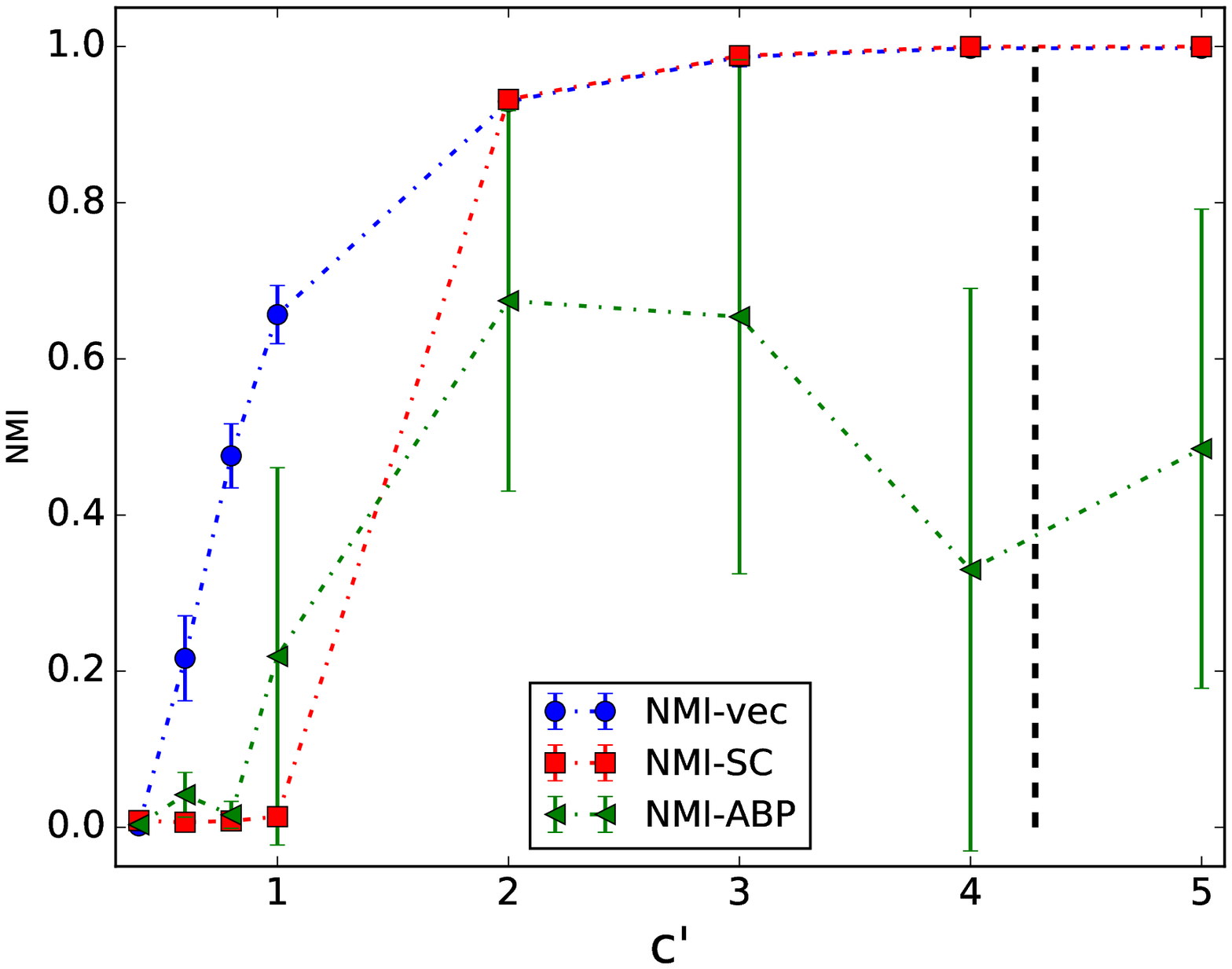}
\includegraphics[width=0.75\linewidth]{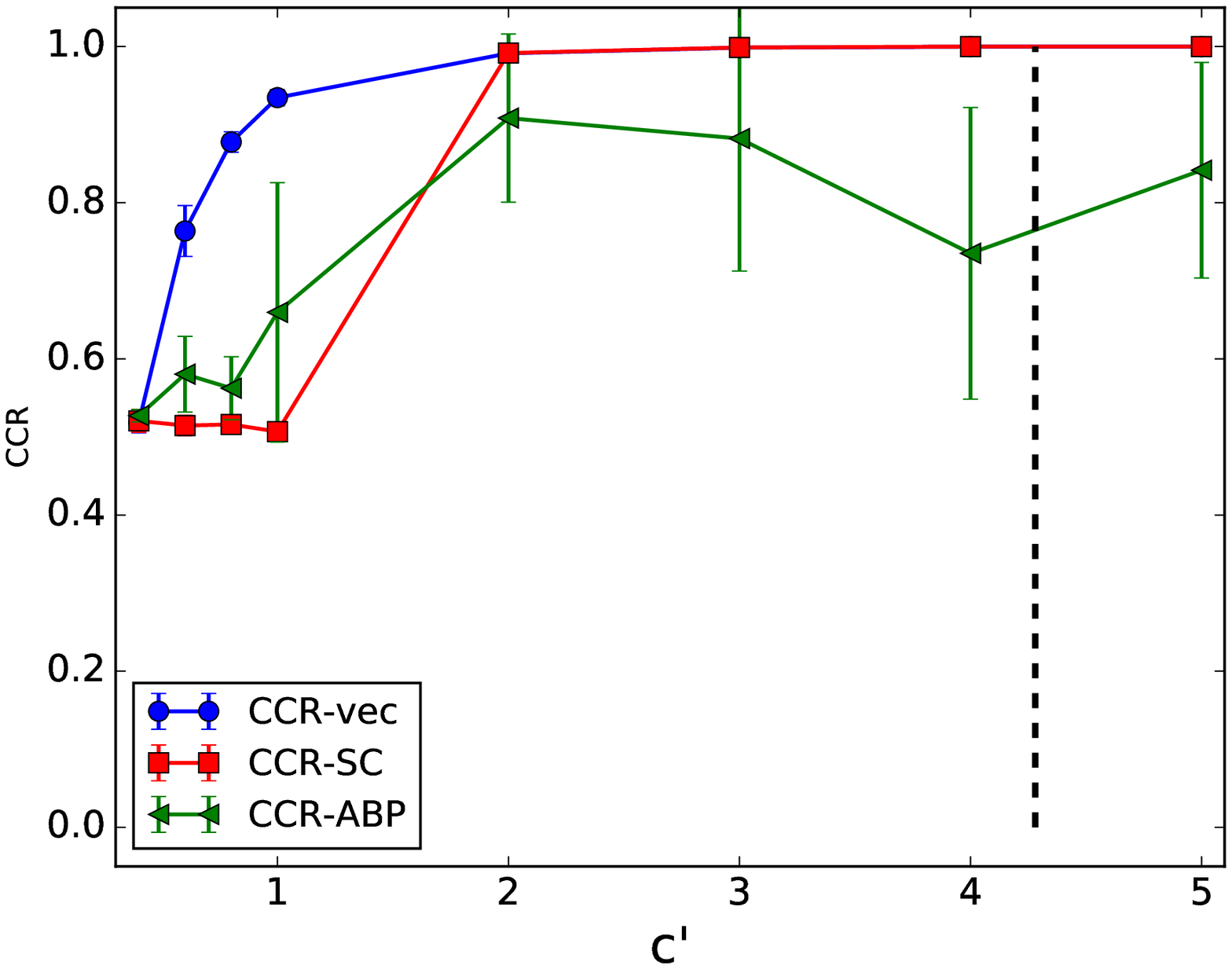}
\caption{\small NMI (\revision{top sub-figure}, dashed curves) and CCR
  (\revision{bottom sub-figure}, solid curves) versus sparsity level
  $c^{\prime}$ for VEC, SC, and ABP on SBM graphs with logarithmic
  scaling. Here, $\mathbf{p}$ is uniform, $K=2,\lambda = 0.9$, and
  $n=10000$. The vertical dashed-line is at $c^{\prime}_{\text{exact}}
  = 4.3$.}
\label{fig:strong_trans}
\end{figure}
%
%
%
We now turn to explore the behavior of VEC near the exact recovery
limit.
Figure~\ref{fig:strong_trans} plots NMI and CCR as a function of
increasing sparsity level $c^{\prime}$ for SBM graphs under {\it
  logarithmic} node degree scaling fixing $K=2$, $N=10000$, and
$\lambda = 0.9$. In this setting, exact recovery is solvable if, and
only if, $c^{\prime} > c^{\prime}_{\text{exact}} \approx 4.3$ ({\it
  cf. Condition 2} in Sec.~\ref{sbm-settings}).
As can be seen in Fig.~\ref{fig:strong_trans}, the $\text{CCR}$ and
$\text{NMI}$ values of VEC converge to $1.0$ as $c^{\prime}$ increases
far beyond $c^{\prime}_{\text{exact}}$. Therefore, VEC empirically
attains the exact recovery limit. We note that SC can match the
performance of VEC when $c^{\prime}$ is large, but cannot correctly
detect communities for very sparse graphs ($c^\prime \leq 1$). Note
also that VEC significantly outperforms ABP in this scaling
scheme. 

%
\begin{figure*}[!htb]
\begin{minipage}[b]{0.33\linewidth}
  \centering
  \centerline{\includegraphics[width=1\linewidth]{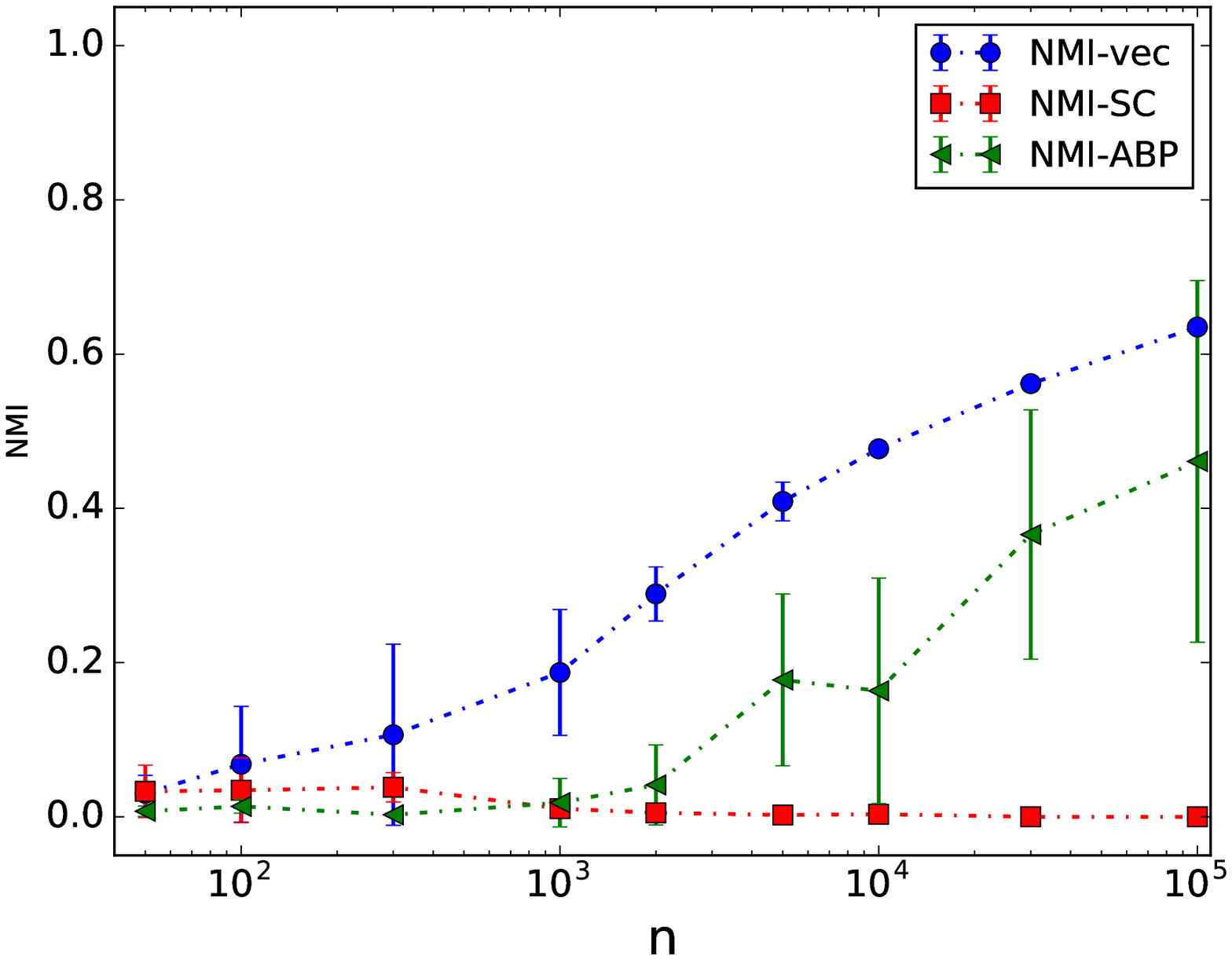}} 
\end{minipage}
\begin{minipage}[b]{0.33\linewidth}
  \centering
  \centerline{\includegraphics[width=1\linewidth]{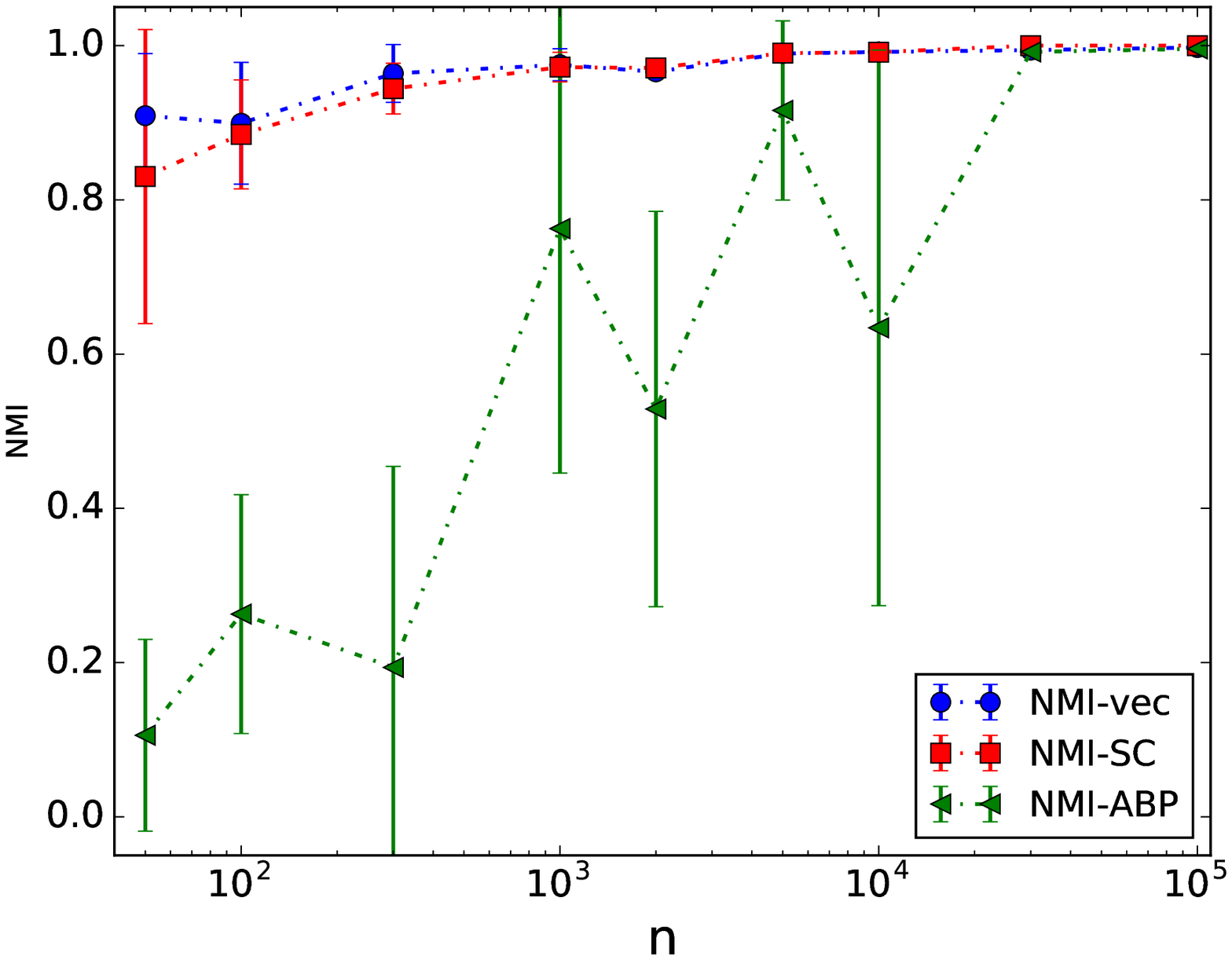}}  
\end{minipage}
\begin{minipage}[b]{0.33\linewidth}
  \centering
  \centerline{\includegraphics[width=1\linewidth]{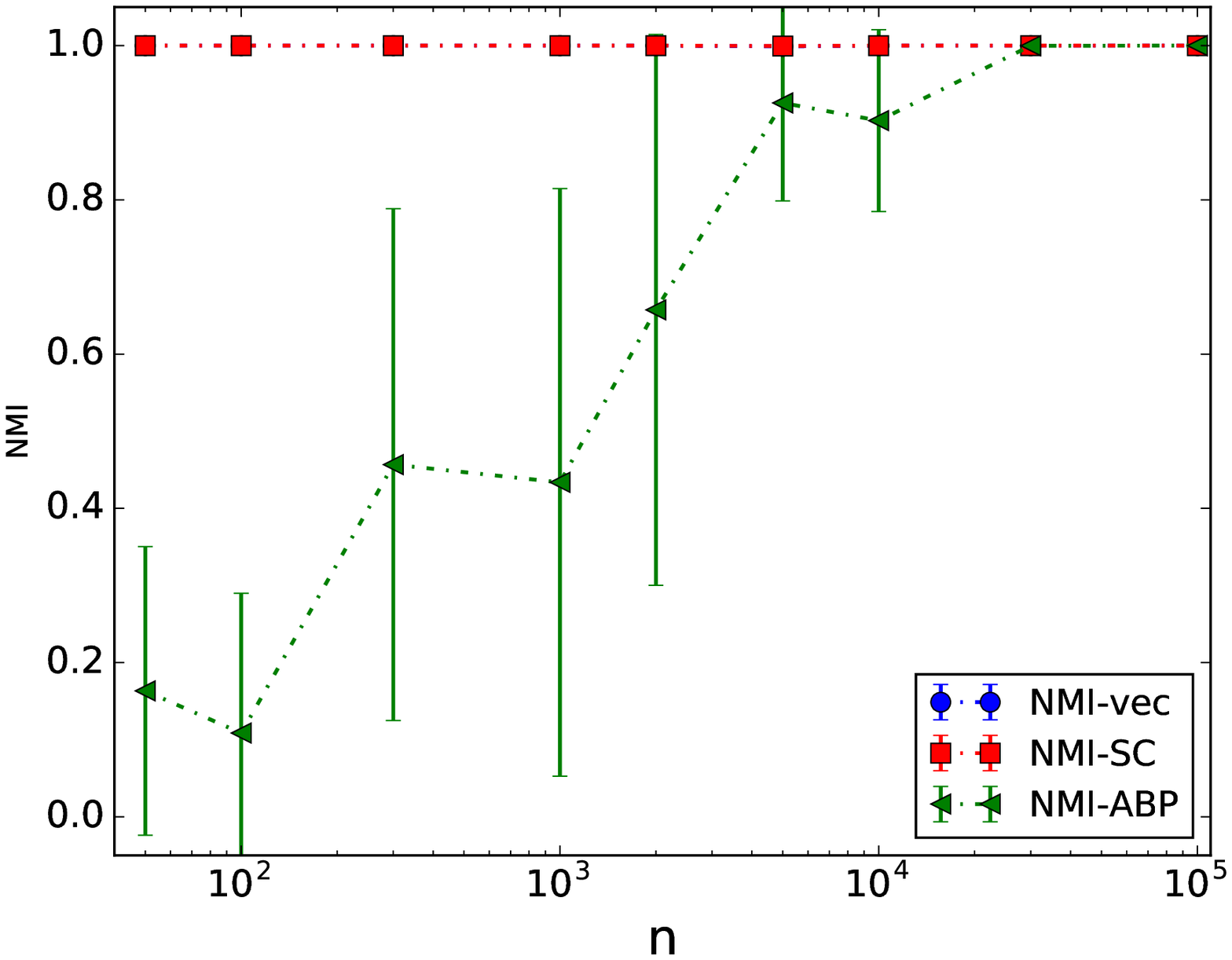}} 
\end{minipage}
\begin{minipage}[b]{0.33\linewidth}
  \centering
  \centerline{\includegraphics[width=1\linewidth]{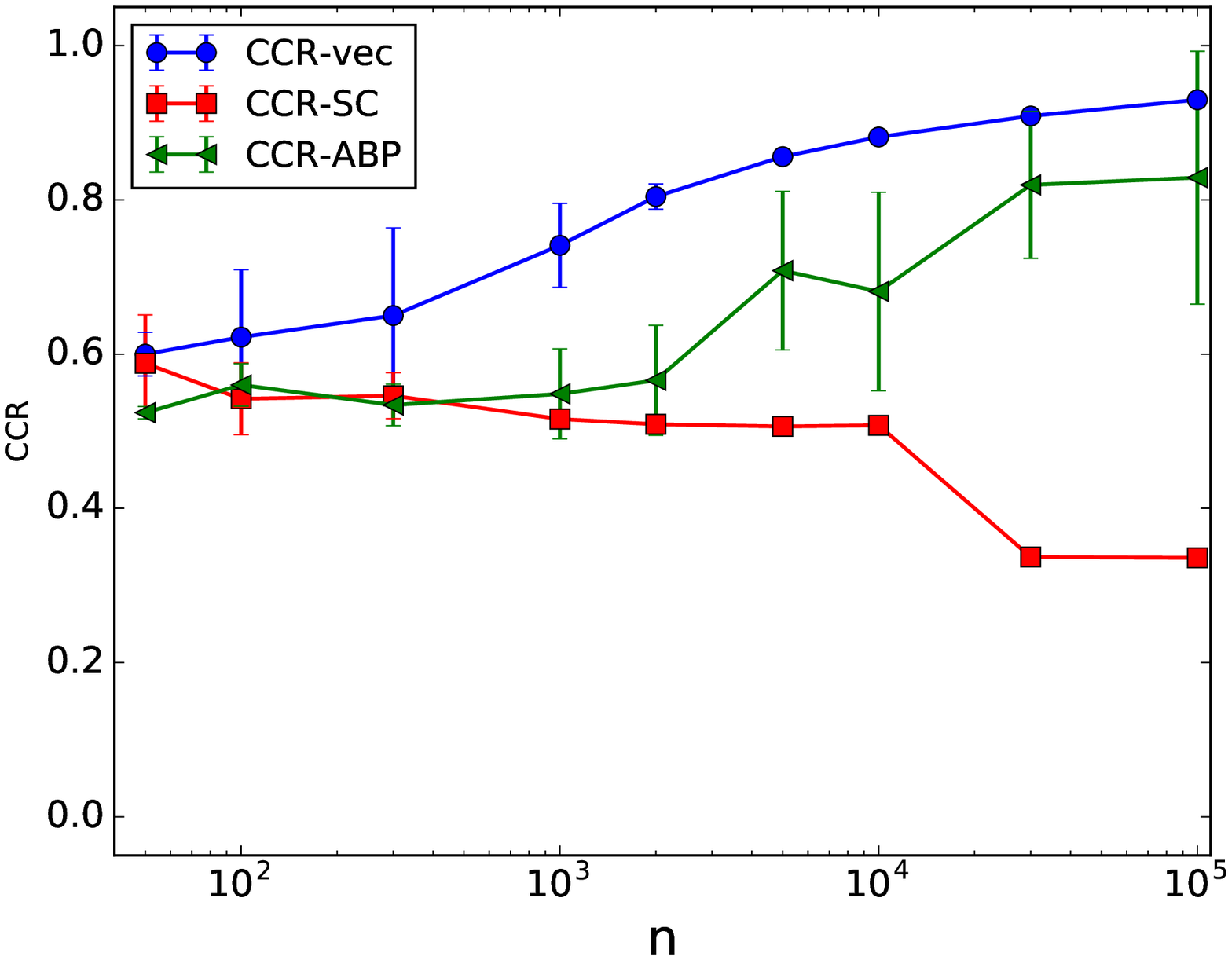}} 
  \centerline{\small (a) $c^{\prime} = 0.6$}\medskip
\end{minipage}
\begin{minipage}[b]{0.33\linewidth}
  \centering
  \centerline{\includegraphics[width=1\linewidth]{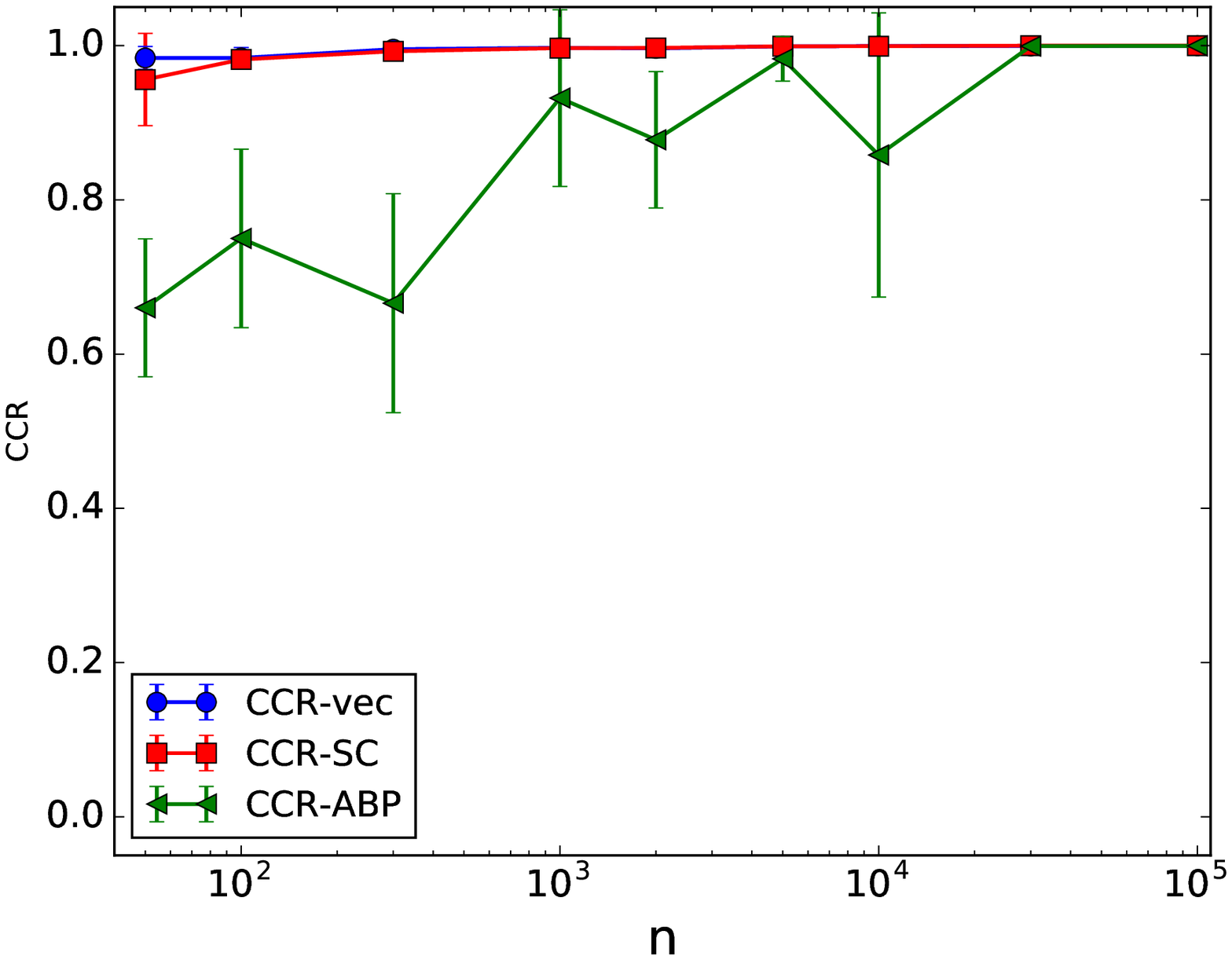}} 
  \centerline{\small (b) $c^{\prime} = 2.5$}\medskip
\end{minipage}
\begin{minipage}[b]{0.33\linewidth}
  \centering
  \centerline{\includegraphics[width=1\linewidth]{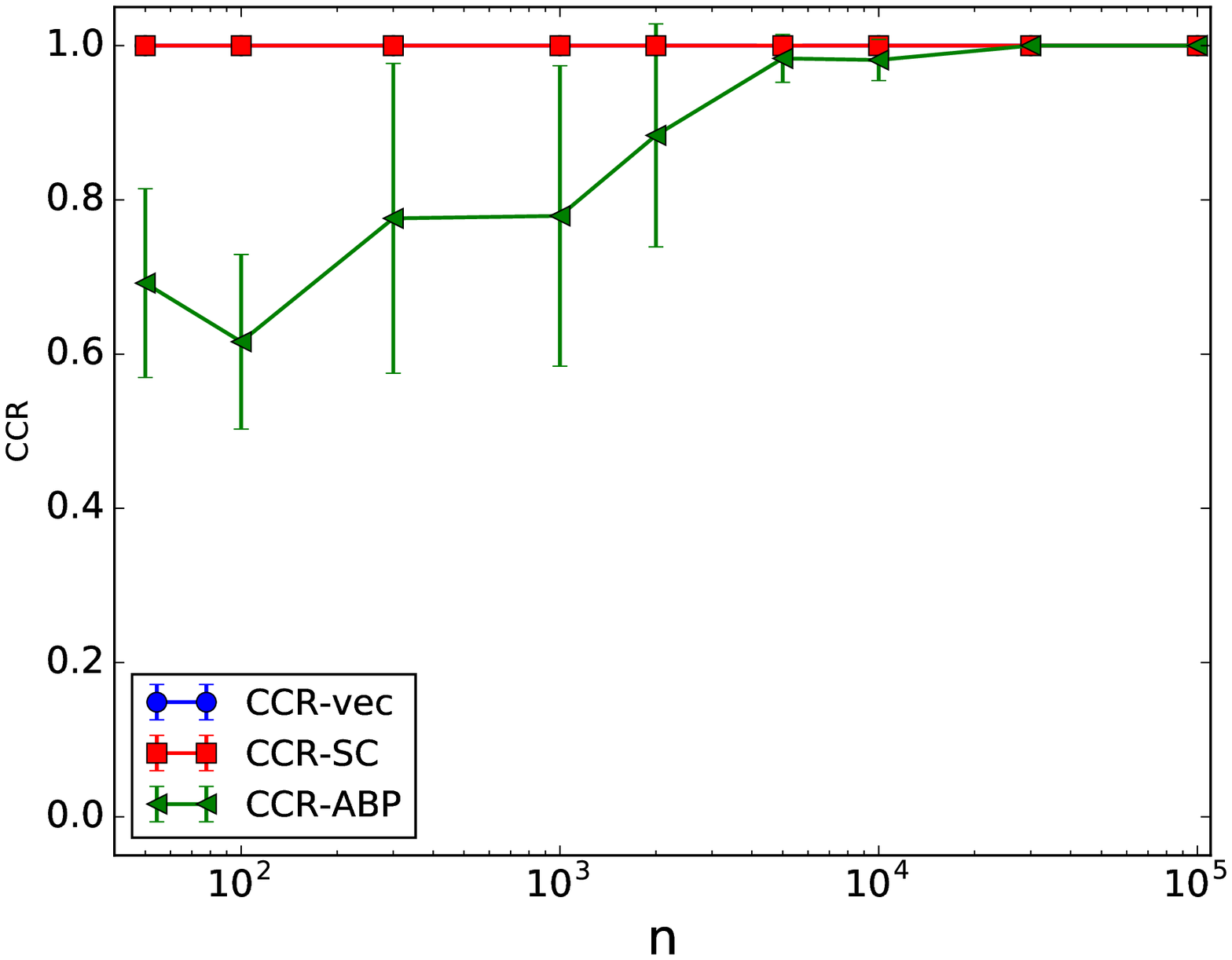}}
  \centerline{\small (c) $c^{\prime} = 4.5$}\medskip
\end{minipage}  
%
%
\caption{\small NMI (top sub-figure, dashed curves) and CCR (bottom
  sub-figure, solid curves) versus number of nodes $n$ for VEC, SC,
  and ABP on SBM graphs with logarithmic scaling. Here, $\mathbf{p}$
  is uniform, $K=2$, and $\lambda=0.9$. In subplots $(a)$ and $(b)$,
  $c^{\prime}<c_{\text{exact}}^{\prime}$ while in subplot $(c)$,
  $c^{\prime} > c_{\text{exact}}^{\prime}$. We note that in subplot
  $(c)$, the curves of VEC and SC are on top of each other since they
  have very similar performance in this setting.}
%
\label{fig:logarithm_scaling}
\end{figure*}
We also compared the behavior of VEC, ABP, and SC algorithms for
increasing graph sizes $n$. We set $K=2, \lambda =0.9$, and
$\mathbf{p} \sim$ uniform. Figure~\ref{fig:logarithm_scaling}
illustrates the performance of VEC, SC, and ABP as a function of the
number of nodes $n$ for three different choices of $c^{\prime}:
c^{\prime} = 0.6, 2.5$, and $4.5$. Since exact recovery requires
$c^{\prime} \geq c^{\prime}_{\text{exact}} \approx 4.3$, only the
third choice of $c^{\prime}$ guarantees exact recovery asymptotically.

As can be seen in Fig.~\ref{fig:logarithm_scaling}, when $c^{\prime}$
is above the exact recovery condition (see
Fig.~\ref{fig:logarithm_scaling} $(c)$), the proposed algorithm VEC
can achieve exact recovery, i.e., $\text{CCR}\approx 1$ and
$\text{NMI}\approx 1$. In this setting, the proposed VEC algorithm can
be observed to achieve exact recovery even when the number of nodes
$n$ is relatively small. On the other hand, when $c^{\prime}$ is below
the exact recovery condition (see Figs.~\ref{fig:logarithm_scaling}
$(a)$ and $(b)$), as $n$ increases, the accuracy of VEC increases and
converges to a value that is somewhere between random guessing
($\text{CCR} = 0.5, \text{NMI} =0.0$) and exact recovery ($\text{CCR}
= 1.0, \text{NMI} = 1.0$).

We note that among the compared baselines, the performance of SC is
similar to that of VEC when $c^{\prime}$ is large (relatively dense
graph) but its performance deteriorates when $c^{\prime}$ is small
(sparse graph). As shown in Fig.~\ref{fig:logarithm_scaling} $(a)$,
when the SBM-synthesized graph is relatively sparse, the performance
of SC is close to a random guess while the performance of VEC and ABP
increases with the number of nodes $n$. This observation is consistent
with known theoretical results \cite{abbe2015,SC2RCY}.

%
%

\subsection{Robustness of proposed approach}
\label{sec:param-sensitive}
\begin{figure*}[htb!]
\begin{minipage}[b]{0.24\linewidth}
  \centering
  \centerline{\includegraphics[width=1\linewidth]{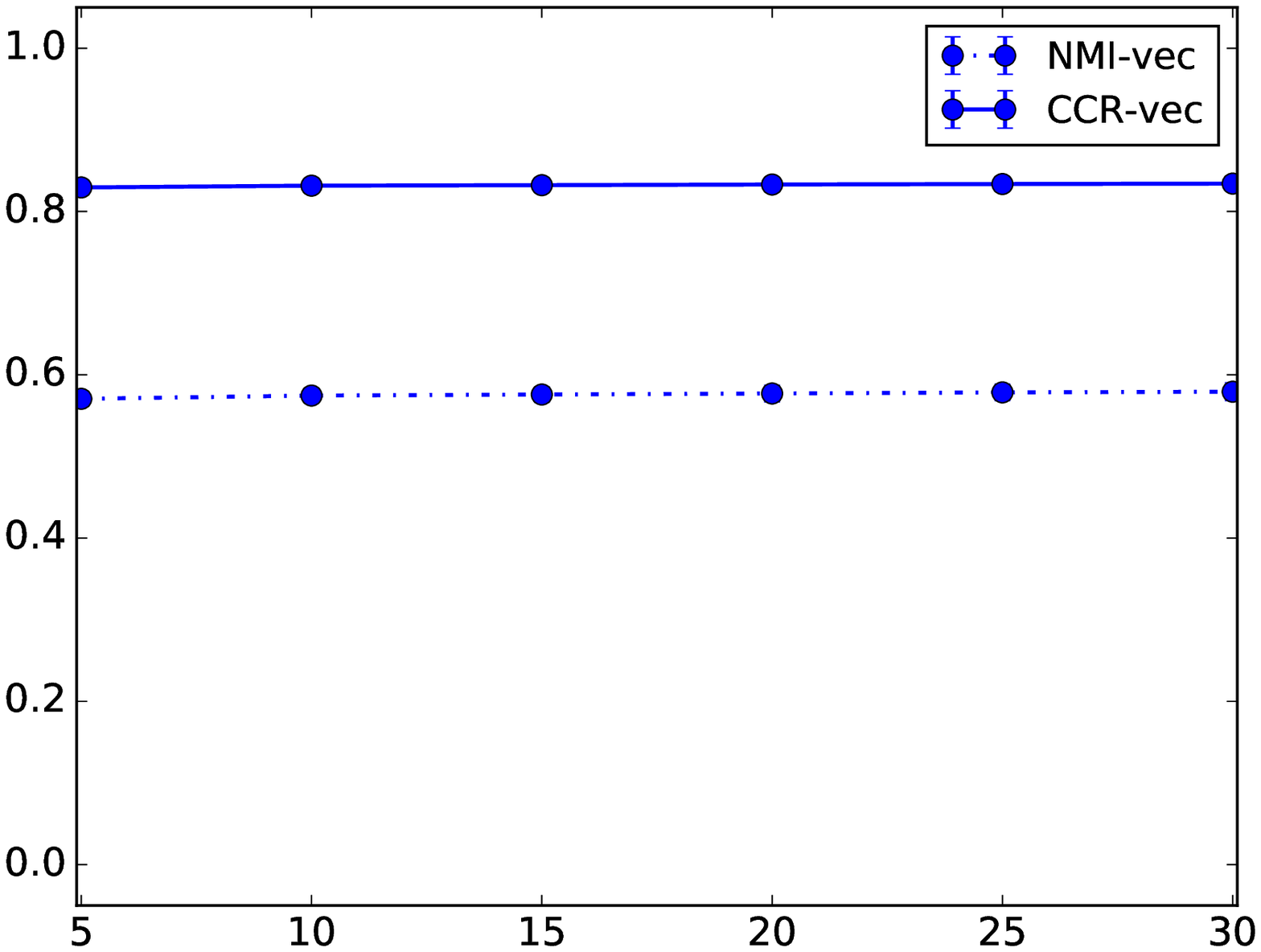}}
%
%
  \centerline{\small (a) $\numpath$}\medskip
\end{minipage}
\begin{minipage}[b]{0.24\linewidth}
  \centering
  \centerline{\includegraphics[width=1\linewidth]{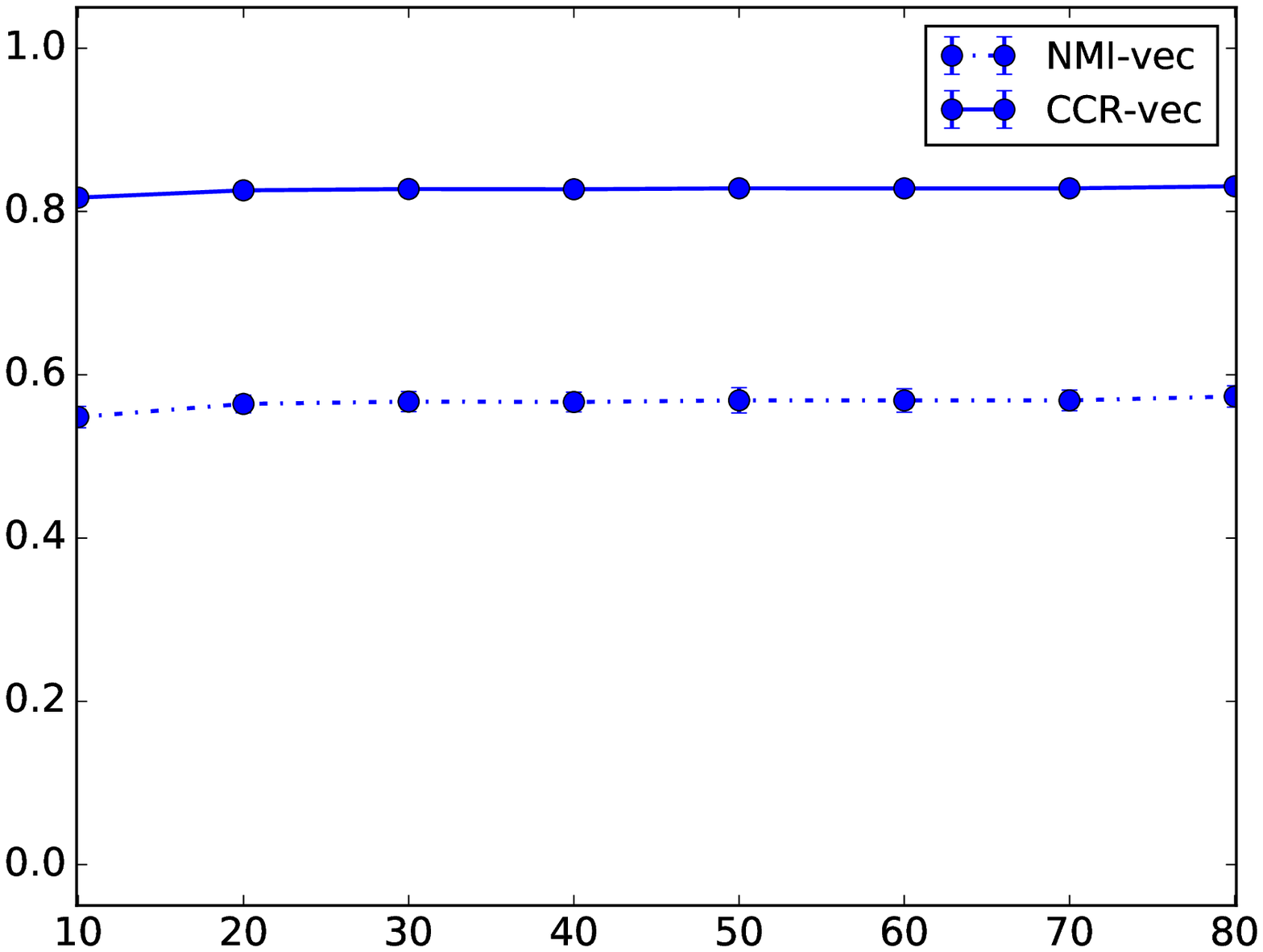}}
%
%
  \centerline{\small (b) $\ell$}\medskip
\end{minipage}
\begin{minipage}[b]{0.24\linewidth}
  \centering
  \centerline{\includegraphics[width=1\linewidth]{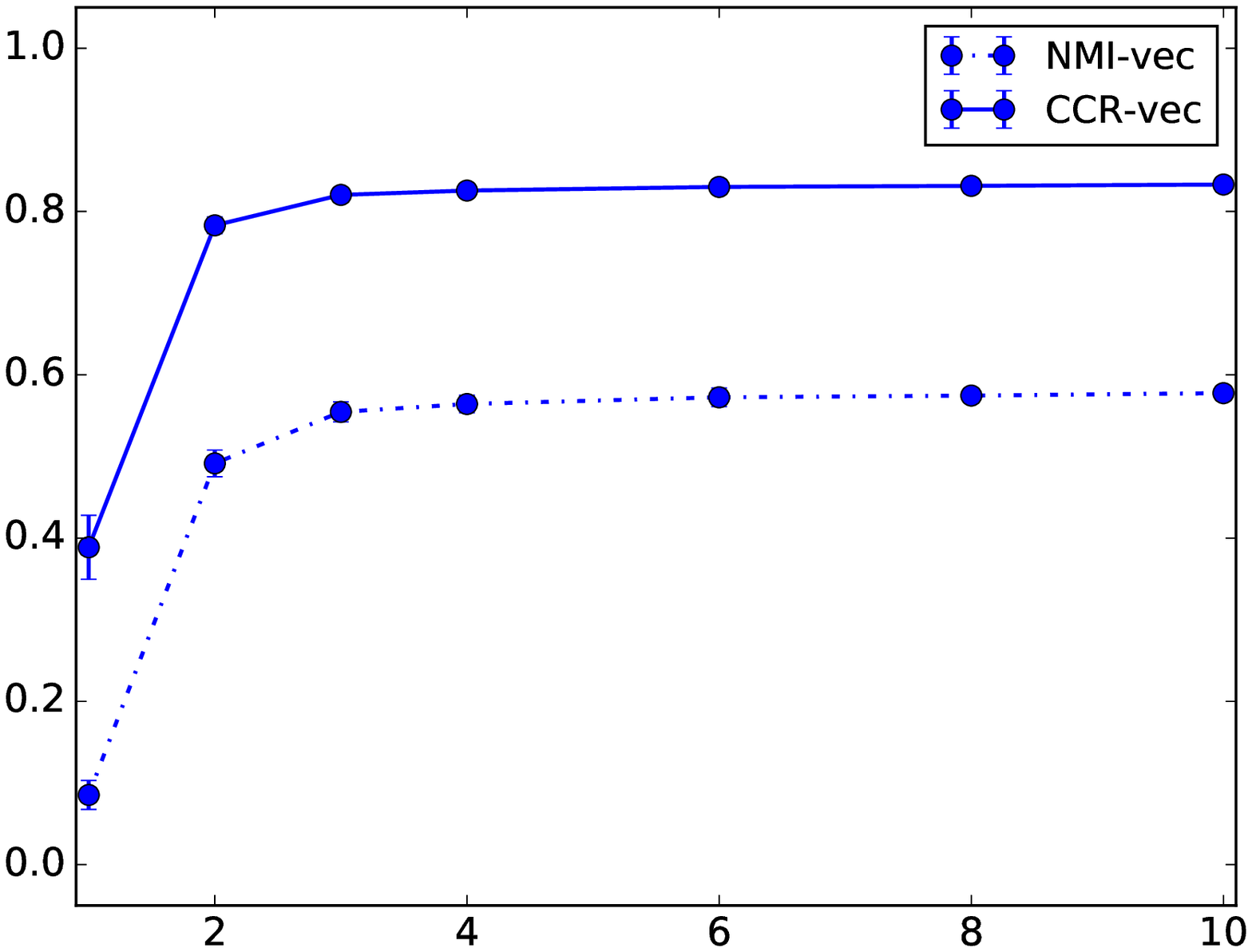}}
%
%
  \centerline{\small (c) $w$}\medskip
\end{minipage}
\begin{minipage}[b]{0.24\linewidth}
  \centering
  \centerline{\includegraphics[width=1\linewidth]{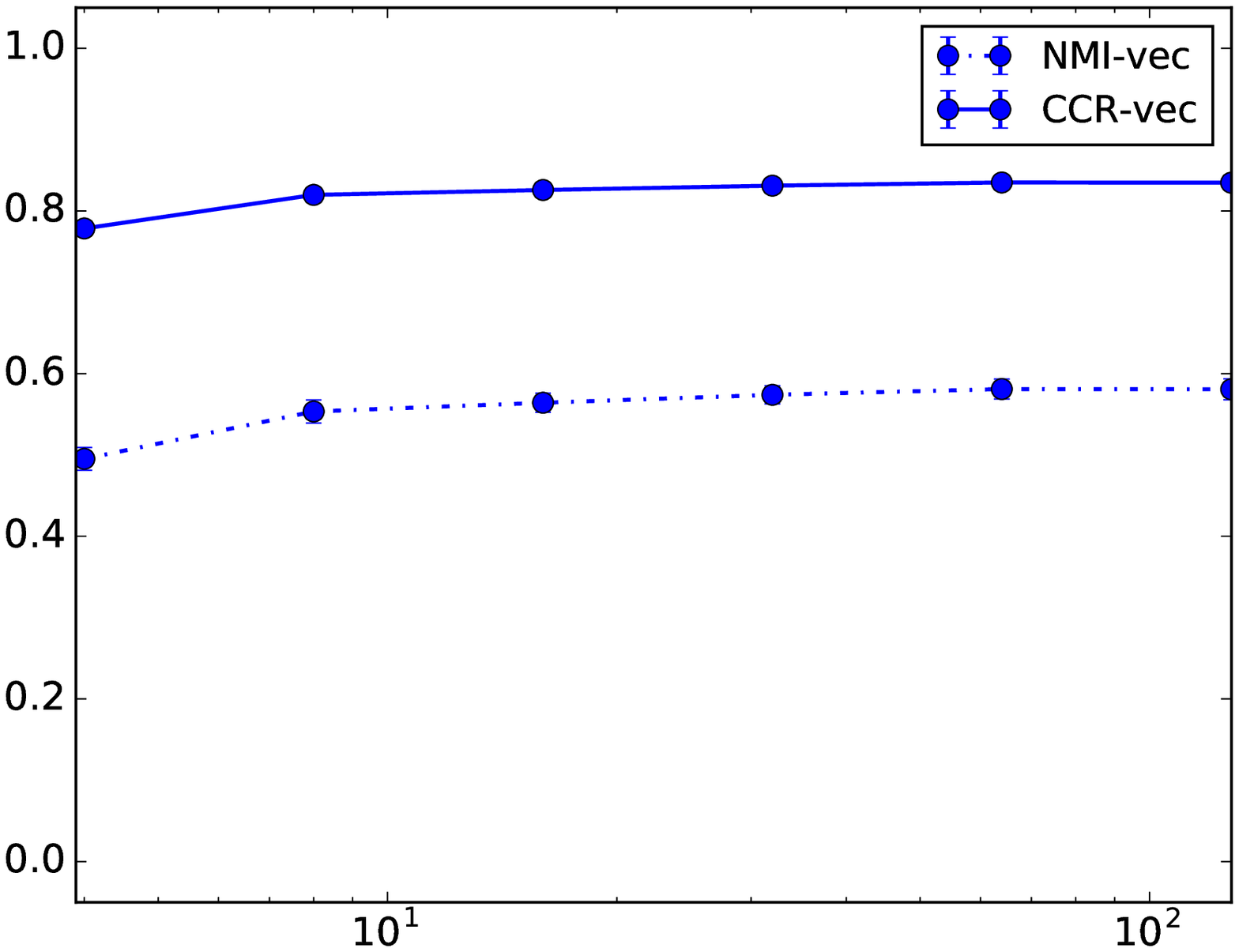}}
%
%
  \centerline{\small (d) $\log(d)$}\medskip
\end{minipage}
%
%
\caption{\small NMI (dashed curves) and CCR (solid curves) as a
  function of different algorithm parameters of VEC on SBM graphs: (a)
  the number of random paths simulated from each node $\numpath$, (b)
  the length of each random path $\ell$, (c) the size of the local
  window $w$, and (d) the embedding dimension $d$. In each subplot,
  only one parameter is varied keeping others fixed. When fixed, the
  default parameter values are $\numpath=10, \ell=60, w=8, d=50$.}
\label{fig:params}
\end{figure*}

\noindent {\bf Parameter sensitivity}: The performance of VEC depends
on the number of random paths per node $\numpath$, the length of each
path $\ell$, the local window size $w$, and the embedding dimension
$d$.
We synthesized SBM graphs under logarithmic scaling with $K=5,
N=10,000, c^{\prime}=2, \lambda = 0.9$ and applied VEC with different
choices for $\numpath$, $\ell$, $w$, and $d$. The results are
summarized in Fig.~\ref{fig:params}.
While the performance of VEC is remarkably insensitive to $\numpath$,
$\ell$, and $d$ across a wide range of values, a relatively large
local window size $w\geq 3$ appears to be essential for attaining good
performance ({\it cf.}  Fig.~\ref{fig:params}(c)).
This suggests that incorporating a larger graph neighborhood is
critical to the success of VEC.

\noindent {\bf Effect of random initialization in VEC and ABP}: We
also studied the effect of random initialization in VEC and ABP. We
synthesized two SBM graphs as described in
Table~\ref{table:random}. For a fixed graph, we run VEC and ABP $10$
times and summarize the mean and standard deviation values of NMI and
CCR. We observe that the variance of ABP is an order of magnitude
higher than VEC indicating its high sensitivity to initialization.
%
%
\begin{table}[htb!]
\caption{\small Means and standard deviations of NMI and CCR for $10$
  runs on the same graph. Sim1: a graph with constant scaling, $K=5,
  N=10000, c=15.0, \lambda =0.9$. Sim2: a graph with logarithmic
  scaling, $K=2, N=10000, c^{\prime}=2.0, \lambda =0.9$.
}
\label{table:random}
\centering
\footnotesize
\begin{tabular}{ |l|c|c|c|c|}
\hline
& \multicolumn{2}{ |c| }{NMI} & \multicolumn{2}{|c|}{CCR} \\
\hline
Expt. & VEC & ABP & VEC & ABP \\ \hline
Sim1 & $0.42\pm 0.004$ & $0.14\pm 0.03$ & $0.74\pm 0.002$ & $0.42\pm
0.06$ \\ \hline
Sim2 & $0.96\pm 0.002$ & $0.73\pm 0.37$ & $0.99\pm 0.0003$ & $0.93\pm
0.15$ \\ \hline
%
%
\end{tabular}
\end{table}

\subsection{Summarizing overall performance via Performance Profiles}
\label{performanceprofiles}
So far we presented and discussed the performance of VEC, ABP, and SC
across a wide range of parameter settings. All results indicate that
VEC matches or outperforms both ABP and SC in almost all scenarios. In
order to summarize and compare of the {\it overall} performance of all
three algorithms across the wide range of parameter settings that we
have considered, we adopt the commonly used Performance Profile
\cite{PP} as a ``global'' evaluation metric. Formally, let
$\mathcal{P}$ denote a set of experiments and $a$ a specific
algorithm. Let $Q(a,e)$ denote the value of a performance metric
attained by an algorithm $a$ in experiment $e$ where higher $Q$ values
correspond to better performance. Then the performance profile of $a$
at $\tau\in[0,1]$ is the fraction of the experiments in which the
performance of $a$ is at least a factor $(1-\tau)$ times as good as
the best performing algorithm in that experiment, i.e.,
\begin{equation}
PP_{a}(\tau) := \frac{|\{e: Q(a,e)\geq (1-\tau)\max_{a'}
  Q(a',e)\}|}{|\mathcal{P}|}
\end{equation}
The Performance Profile is thus an empirical cumulative distribution
function of an algorithm's performance relative to the best-performing
algorithm in each experiment.  We calculate $PP_{a}(\tau)$ for $\tau
\in (0,1)$. The higher a curve corresponding to an algorithm, the more
often it outperforms the other algorithms.

For simplicity, we {\it only} consider the simulation settings for the
{\it planted partition model} SBMs in the constant degree scaling
regime. We set $c\in \{ 2, 5, 10, 15 \}$, $K\in \{2, 5, 10\}$, and $N
\in\{ 1e2, 1e3, 1e4, 1e5 \}$. For each combination of settings
$(c,K,N)$, we conduct $5$ independent random repetitions of the
experiment. Thus overall the Performance Profile is calculated based
on $240$ experiments.

\begin{figure}[htb!]
\centering
\includegraphics[width=0.75\linewidth]{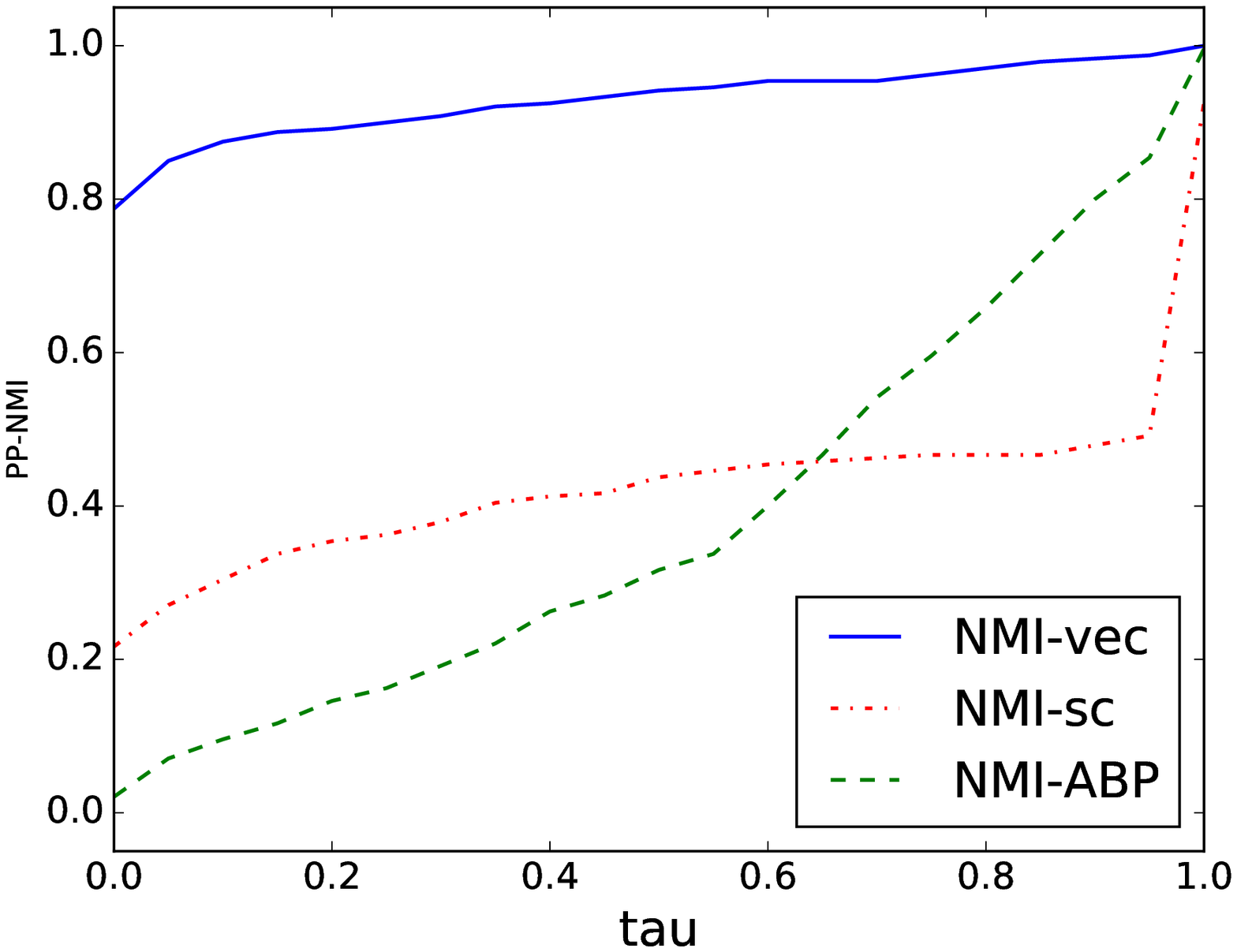}
\includegraphics[width=0.75\linewidth]{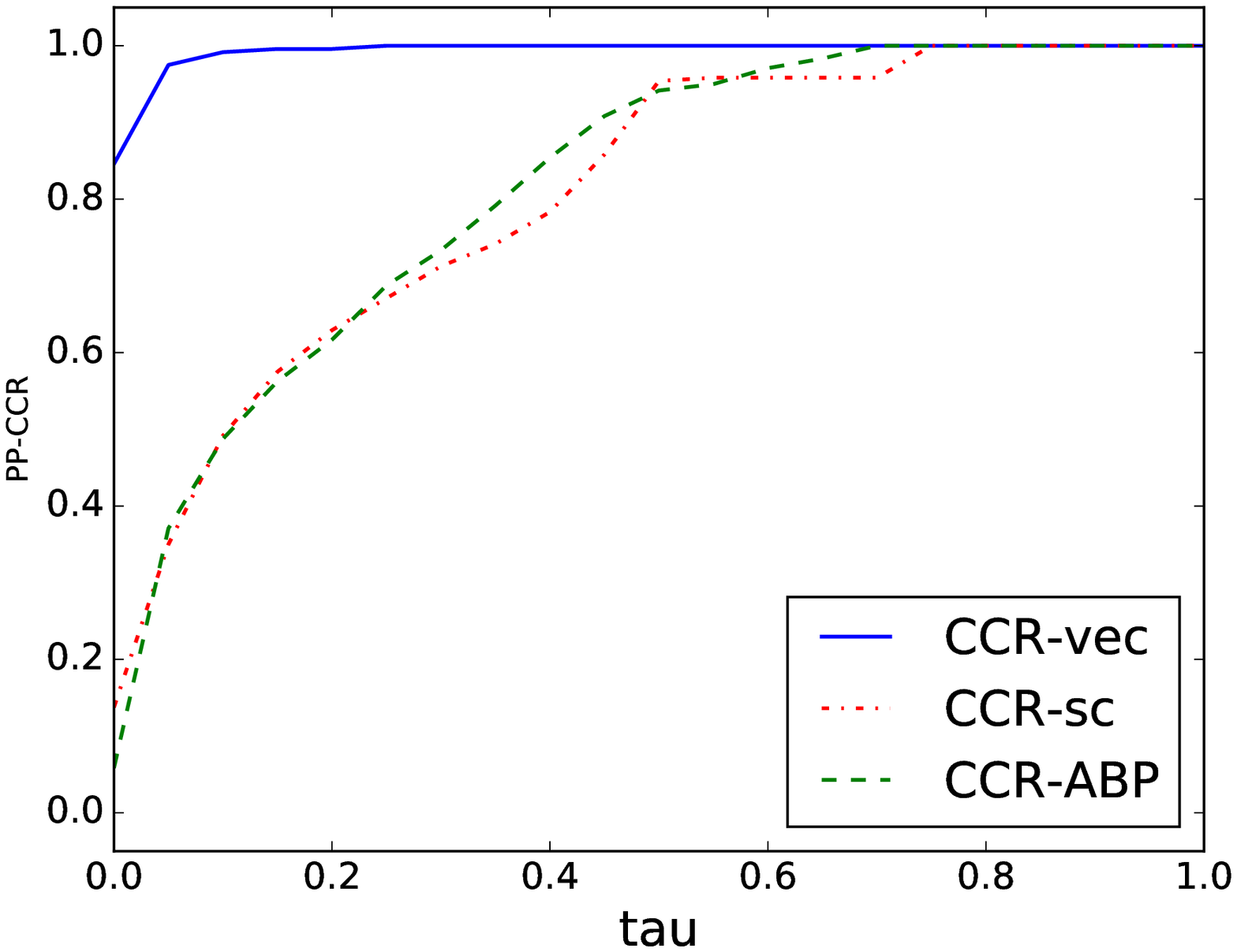}
%
\caption{\small NMI (top sub-figure) and CCR (bottom sub-figure)
  Performance Profiles for VEC, SC, and ABP on SBM graphs with
  constant degree scaling. Here, $c\in \{ 2, 5, 10, 15 \}$, $K\in \{2,
  5, 10\}$, $N \in\{ 1e2, 1e3, 1e4, 1e5 \}$, and $5$ random runs for
  each combination of settings for a total of $240$ distinct
  experiments.}
%
%
\label{fig:PP}
\end{figure}

Figure~\ref{fig:PP} shows the performance profiles for both NMI and
CCR metrics. From the figure it is clear that VEC dominates both ABP
and SC and that ABP and SC have similar performance across many
experiments.

\subsection{\revision{Degree-corrected SBM}}
\label{dc-result}
\revision{Our experiments thus far focused on the SBM where the
  expected degree is constant across all nodes within the same
  community (degree homogeneity). In order to compare the robustness
  of different algorithms to degree {\it heterogeneity}, we considered
  the degree-corrected SBM \cite{karrer2011stochastic} (DC-SBM), which
  generates edge-weighted graphs with a power law within-community
  degree distribution that is observed in many real-world graphs. We
  adopted the following generative procedure which was proposed in
  \cite{karrer2011stochastic}: (1) Assign the latent community labels
  using the same procedure as in the SBM; (2) Within each community,
  sample a parameter $\theta_i$ from a power law distribution for each
  node $i$ within this community, and normalize these $\theta_i$'s so
  that they sum up to $1$; (3) For any two nodes $i$ and $j$, sample a
  weighted edge $\{i,j\}$ with weights drawn from a Poisson
  distribution with mean $\theta_i\theta_jQ_n(\pi_i,\pi_j)$. We
  omitted the steps of generating self-loops since they will be
  ignored by the community detection algorithms. Our proposed
  algorithm VEC can be modified to handle weighted graphs by setting
  the random walk transition probabilities to be proportional to the
  edge weights. The ABP and SC algorithms can also be suitably adapted
  to work with weighted graphs.}

\revision{We simulated graphs with $n = 1000, \lambda = 0.9$ and the
  power law distribution for $\theta_i$ with power $-2.5$ and a
  minimum $\theta_i$ value of $10$. We considered a number of sparsity
  settings used to simulate SBM graphs in previous sections. For each
  setting, the mean NMI and CCR values and their confidence intervals
  (based on $500$ random graphs ) are summarized in
  Table~\ref{table:dc-sbm}. These results indicate that our algorithm
  still outperforms the competing approaches. We would like to point
  out that even though the DC-SBM and SBM parameters are similar, the
  similarity of parameter settings does not imply similarity of
  information-theoretic limits. To the best of our knowledge, the
  information-theoretic limits of recovery for the DC-SBM is still
  open.}
\begin{table*}[htb!]
\caption{\small \revision{Results for degree-corrected SBM graphs.} }
\label{table:dc-sbm}
\centering{
\footnotesize
\begin{tabular}{ |l|c|c|c|c|c|c|}
\hline
Setting& \multicolumn{3}{ |c| }{NMI} & \multicolumn{3}{|c|}{{CCR}}
\\ \hline
 & VEC & SC & ABP  & VEC & SC & ABP  \\ \hline
$K=2, c=10$, constant scaling & $0.610\pm 0.057$ & $0.007\pm 0.01$ & $0.008\pm0.05$ & $0.920\pm0.02$ & $0.512\pm0.01$ & $0.519\pm 0.04$
\\ \hline
$K=5, c=10$, constant scaling & $0.167\pm 0.02$ & $0.022\pm 0.01$ & $0.006\pm 0.01$ & $0.418\pm 0.03$ & $0.224\pm 0.001$ & $0.227\pm 0.02$
\\ \hline
$K=2, c^{\prime}=2$, log scaling & $0.737\pm 0.07$ & $0.018\pm 0.04$ & $0.014\pm 0.07$ & $0.953\pm 0.02$ & $0.515\pm 0.02$ & $0.522\pm 0.04$
\\ \hline
$K=5, c^{\prime}=2$, log scaling & $0.295\pm 0.05$ & $0.018\pm 0.01$ & $0.007\pm0.01$ & $0.573\pm 0.07$ & $0.224\pm0.01$ & $0.226\pm0.02$
\\ \hline
\end{tabular}
}
\end{table*}

\section{Experiments with Real-World Graphs}
%
%
Having comprehensively studied the empirical performance of VEC, ABP,
and SC on SBM-based synthetic graphs, in this section we turn our
attention to real-world datasets that have ground truth
(non-overlapping) community labels. Here we use only NMI\cite{NMI} to
measure the performance. 
%

We consider two benchmark real-world graphs: the {\it Political Blogs}
network \cite{blogdataset} and the {\it Amazon} co-purchasing network
\cite{snapnets}.
Since the original graphs are directed, we convert it to undirected
graphs by forming an edge between two nodes if {\it either} direction
is part of the original graph.
The basic statistics of the datasets are summarized in
Table~\ref{table:realdatastats}.  Here, $\hat{\lambda}$ and
$\hat{c^\prime}$ are the maximum likelihood estimates of $\lambda$ and
$c^\prime$ respectively in the {\it planted partition model} SBM under
logarithmic scaling.
Note that in {\it Amazon}, the ground truth community proportions are
highly unbalanced.
\begin{table}[hbt!]
\centering
\caption{\small Summary of real-world dataset parameters.}
\small
\label{table:realdatastats}
\small
\begin{tabular}{|l|c|c|c||c|c|c|}
\hline
{\bf Dataset} & $n$ & $K$ & 
\# edges & $\hat{\lambda}$ & 
$\hat{c^\prime}$ & $\max_{k} \hat{p}_{k}$
\\ \hline
{\it Blogs} & $1,222$ & $2$ & $16,714$ & $0.89$ & $6.9$ & $0.52$
\\ \hline
{\it Amazon} & $334,844$ & $4$ & $925,803$ & $0.94$ & $0.7$ & $0.74$
\\ \hline
\end{tabular}
\end{table}

\begin{figure}[htb!]
\centering
\includegraphics[width=0.75\linewidth]{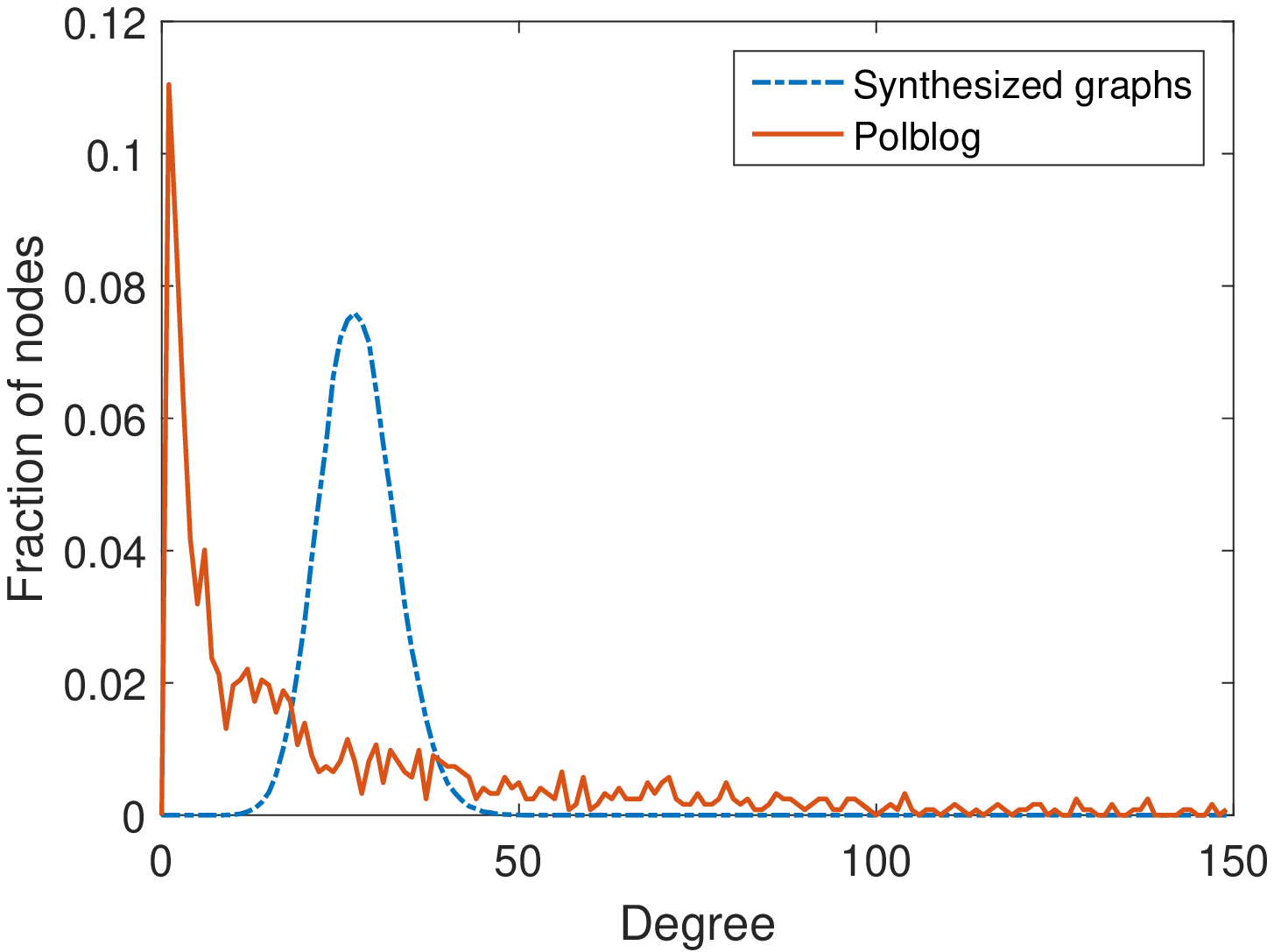}
\includegraphics[width=0.75\linewidth]{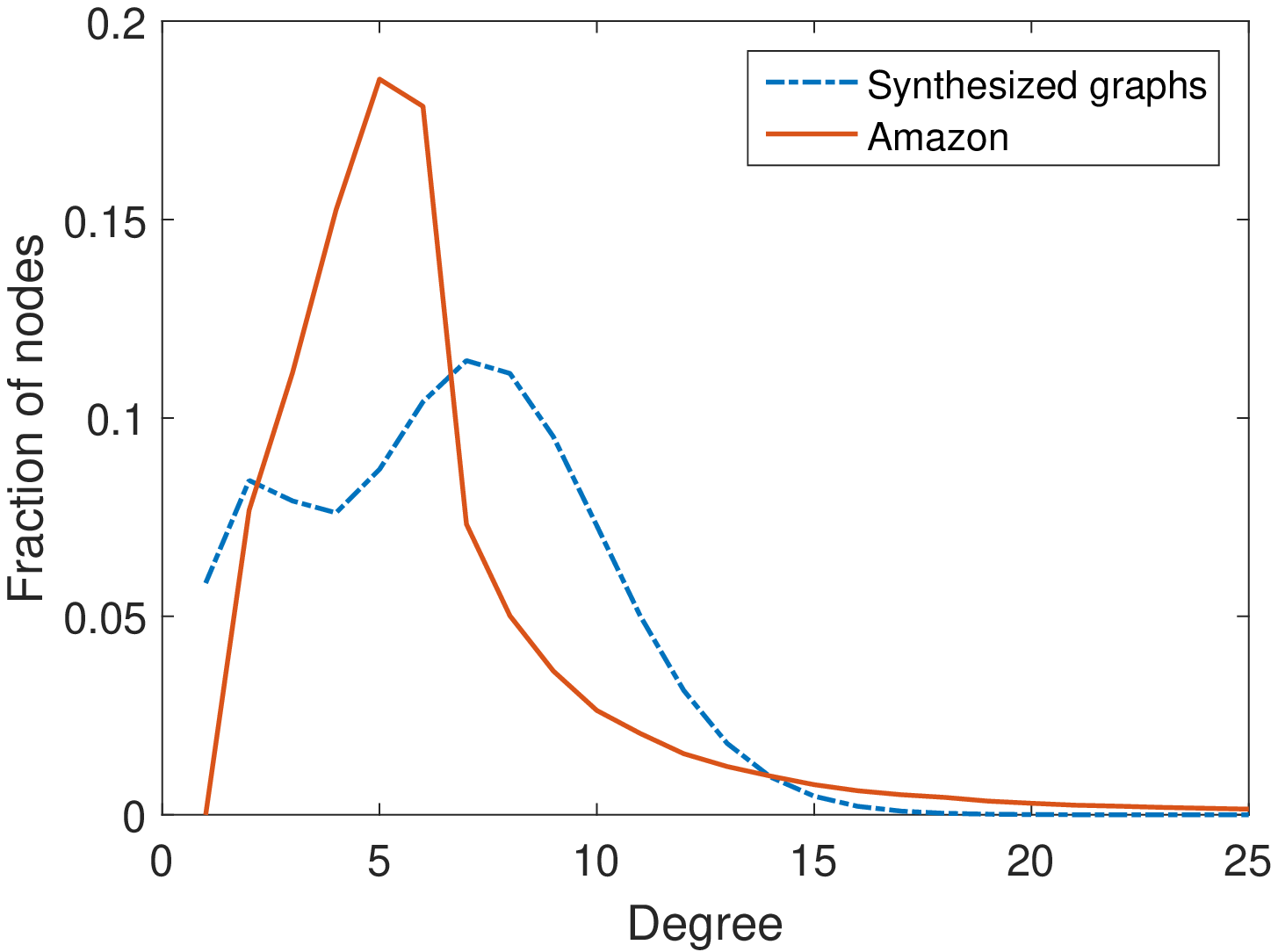}
%
\caption{\small \revision{Empirical degree distributions of real and
    synthesized graphs (averaged over 200 trials) for the {\it
      Political Blogs} (top) and {\it Amazon} (bottom) datasets. The
    error bars are too small to be visible in the plots and have been
    omitted. The distributions have long tails and are truncated in
    the figure for better of visualization.}}
\label{fig:degdistb}
\end{figure}

We report NMI and \revision{CCR}
values for VEC, SC, and ABP applied to these datasets. 
%
To apply ABP, we set the algorithm parameters using the fitted SBM
parameters as suggested in \cite{abbe2016nips}.
As shown in Table~\ref{table:realresult}, VEC achieves better accuracy
compared to SC and ABP. 
\begin{table}[htb!]
\caption{\small Results on real-world datasets. }
\label{table:realresult}
\centering{
\footnotesize
\begin{tabular}{ |l|c|c|c|c|c|c|}
\hline
& \multicolumn{3}{ |c| }{NMI} & \multicolumn{3}{|c|}{\revision{CCR}}
\\ \hline
Data & VEC & SC & ABP  & VEC & SC & ABP  \\ \hline
{\it Blogs} & $0.745$ & $0.002$ & $0.686$ & $0.954$ & $0.529$ & $0.925$
\\ \hline
{\it Amazon} & $0.310$ & $0.006$ &$0.025$ & $0.783$ &$0.742$ & $0.762$ \\ \hline
\end{tabular}
}
\end{table}
The performance of SC is noticeably poorer (in terms of both NMI and
\revision{CCR}) compared to both VEC and ABP.
%
%
Interestingly, the NMI of SC on random graphs that are synthetically
generated according to a {\it planted partition} SBM model that best
fits (in a maximum-likelihood sense) the {\it Political Blogs} graph
is surprisingly good: NMI = $1.0$ (average NMI across 10 random
graphs).
This suggests that real-world graphs such as {\it Political Blogs}
have additional characteristics that are not well-captured by a {\it
  planted partition} SBM model. This is further confirmed by the plots
of empirical degree distributions of nodes in real-world and
synthesized graphs in Fig.~\ref{fig:degdistb}. The plots show that the
node degree distributions are quite different in real-world and
synthesized graphs even if the SBM model which is used to generate the
graphs is fitted in a maximum-likelihood sense to real-world
graphs. These results also suggest that the performance of SC is
sensitive to model-mismatch and its good performance on synthetically
generated graphs based on SBMs may not be indicative of good
performance on matching real-world graphs. \revision{In contrast to
  SC,} both VEC and ABP do not seem to suffer from this limitation.

\begin{figure}[htb!]
\centering
\includegraphics[width=0.75\linewidth]{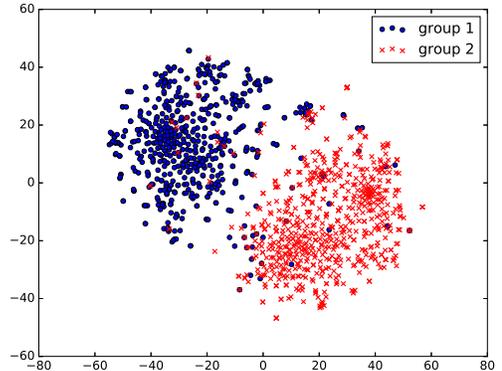} 	
\caption{\small t-SNE visualization of learned embedding vectors in
  the {\it Blogs} dataset. The markers reflect ground-truth groups.}
\label{fig:tsne}
\end{figure}

Finally, we also visualize the learned embeddings in {\it Political
  Blogs} using the now-popular t-Distributed Stochastic Neighbor
Embedding (t-SNE) tool \cite{tSNE} in Fig.~\ref{fig:tsne}. The picture
is consistent with the intuition that nodes from the same community
are close to each other in the latent embedding space.
\label{subsec:real}


%
%

\section{Concluding Remarks} In this work we put forth a novel
framework for community discovery in graphs based on node
embeddings. We did this by first constructing, via random walks in the
graph, a document made up of sentences of node-paths and then applying
a well-known neural word embedding algorithm to it. We then conducted
a comprehensive empirical study of community recovery performance on
both simulated and real-world graph datasets and demonstrated the
effectiveness and robustness of the proposed approach over two
state-of-the-art alternatives. In particular, the new method is able to
attain the information-theoretic limits for recovery in stochastic
block models.

There are a number of aspects of the community recovery problem that
we have not explored in this work, but which merit further
investigation. First, we have focused on undirected graphs, but our
algorithm can be applied `as-is' to directed graphs as well. We have
assumed knowledge of the number of communities $K$, but the node
embedding part of the algorithm itself does not make use of this
information. In principle, we can apply any $K$-agnostic clustering
algorithm to the node embeddings. We have focused on non-overlapping
community detection. It is certainly possible to convert an
overlapping community detection problem with $K$ communities into a
non-overlapping community detection problem with $2^K$ communities,
but this approach is unlikely to work well in practice if $K$ is
large. An alternative approach is to combine the node embeddings with
topic models to produce a ``soft'' clustering. Finally, this study was
purely empirical in nature. Establishing theoretical performance
guarantees that can explain the excellent performance of our algorithm
is an important task which seems challenging at this
time. \revision{One difficulty is the nonconvex objective function of
  the word2vec algorithm. This can be partially addressed by
  constructing a suitable convex relaxation and analyzing its limiting
  behavior (under suitable scaling) as the length of the random walk
  goes to infinity. The limiting objective function will still be
  random since it depends on the observed realization of the random
  graph. One could then examine if the limiting objective, when
  suitably normalized, concentrates around its mean value as the graph
  size goes to infinity.}

\bibliographystyle{IEEEtran}
\bibliography{IEEEabrv,refs4}

\end{document}